\documentclass[a4paper,11pt]{article}
\pdfoutput=1 

\usepackage{jheppub} 

\usepackage{amsmath,amsfonts,amssymb,enumitem}
\usepackage{bbm}
\usepackage{mathrsfs}
\usepackage{comment}
\usepackage{graphicx}
\usepackage[english]{babel} 
\usepackage{url} 
\usepackage{color}
\usepackage{bbold}
\usepackage{physics,amsmath}
\usepackage{adjustbox}
\usepackage[utf8]{inputenc} 
\usepackage[colorlinks=true,citecolor=blue,urlcolor=blue]{hyperref}
\usepackage{bm}
\usepackage{gensymb}
\usepackage{tikz}
\usetikzlibrary{decorations.pathmorphing}
\tikzset{snake it/.style={decorate, decoration=snake}}
\usepackage{microtype}
\usepackage{graphicx}
\usepackage{subcaption}

\renewcommand{\vec}[1]{\bm{#1}}

\newcommand{\be}{\begin{equation}}
\newcommand{\ee}{\end{equation}}
\newcommand{\bes}{\begin{equation*}}
\newcommand{\ees}{\end{equation*}}

\title{\boldmath The Migdal Effect in Semiconductors for Dark Matter with Masses below $\sim \,$100~MeV} 

\author[a]{Kim V. Berghaus,}
\author[b,c,d]{Angelo Esposito,}
\author[a]{Rouven Essig,}
\author[e]{and Mukul Sholapurkar}

\affiliation[a]{C.N. Yang Institute for Theoretical Physics, Stony Brook University, NY 11794, USA}
\affiliation[b]{School of Natural Sciences, Institute for Advanced Study, Princeton, NJ 08540, USA}
\affiliation[c]{Dipartimento di Fisica, Sapienza Universit\`a di Roma, Piazzale Aldo Moro 2, I-00185 Rome, Italy}
\affiliation[d]{INFN Sezione di Roma, Piazzale Aldo Moro 2, I-00185 Rome, Italy}
\affiliation[e]{Department of Physics, University of California, San Diego, CA 92093, USA}

\emailAdd{kim.berghaus@stonybrook.edu}
\emailAdd{angelo.esposito@uniroma1.it}
\emailAdd{rouven.essig@stonybrook.edu}
\emailAdd{msholapurkar@physics.ucsd.edu}

\abstract{
Dark matter scattering off a nucleus has a small probability of inducing an observable ionization through the inelastic excitation of an electron, called the Migdal effect. We use an effective field theory to extend the computation of the Migdal effect in semiconductors to regions of small momentum transfer to the nucleus, where the final state of the nucleus is no longer well described by a plane wave.
Our analytical result 
can be fully quantified by the measurable dynamic structure factor of the semiconductor, which accounts for the vibrational degrees of freedom (phonons) in a crystal.   
We show that, due to the sum rules obeyed by the structure factor, the inclusive Migdal rate and the shape of the electron recoil spectrum is well captured by approximating the nuclei in the crystal as free ions; however, 
the exclusive differential rate with respect to energy depositions to the crystal depends on the phonon dynamics encoded in the dynamic structure function of the specific material.  
Our results now allow the Migdal effect in semiconductors to be evaluated even for the lightest dark matter candidates ($m_\chi \gtrsim 1$~MeV) that can kinematically excite electrons.

}

\date{\today}

\begin{document}
\maketitle

\section{Introduction}
\label{sec:intro}
Dark matter (DM) direct detection experiments seek evidence for non-gravitational interactions between DM particles and ordinary matter. 
In particular, the search for DM interactions with nuclei has seen tremendous progress in the past few decades in probing DM masses above the proton, with  
one- to multi-ton scale detectors now leading the search~\cite{XENON:2018voc,LZ:2022ufs,PandaX-4T:2021bab}. 
In this mass range, typical searches look for DM particles scattering elastically off the nuclei in the detector, as the resulting nuclear recoil energies are sufficiently large to be above detector thresholds.  However, for DM with masses below the GeV-scale, the small recoil energies from elastic DM--nucleus scatterings is challenging to detect, leading to much weaker constraints when compared to those for heavier DM~\cite{SuperCDMS:2020aus,Abdelhameed:2019hmk}.  
Instead, stronger constraints on DM--nucleus interactions can be obtained by searching for inelastic DM processes, in which the DM can transfer an $\mathcal{O}(1)$ fraction of its kinetic energy to the target~\cite{Essig:2011nj,Kouvaris:2016afs,Ibe:2017yqa}. 

For DM--nucleus scattering, one can make use of the Migdal effect, in which the small nuclear recoil is accompanied by a prompt electron, which can absorb most of the DM kinetic energy~\cite{Migdal1939,Migdal:1941, doi:10.1063/1.2995000,Vergados:2004bm,Moustakidis:2005gx,Bernabei:2007jz,Ibe:2017yqa}. 
The first experimental results are promising~\cite{LUX:2018akb,Liu:2019kzq,Essig:2019xkx,EDELWEISS:2019vjv,Aprile:2019jmx,SENSEI:2020dpa,Knapen:2021bwg,COSINE-100:2021poy,SuperCDMS:2022kgp,EDELWEISS:2022ktt,DarkSide:2022dhx} and there is a clear path for improvement~\cite{Essig:2022dfa}. In addition, proposals exist to discover and calibrate the Migdal effect in the laboratory with neutrons~\cite{Nakamura:2020kex,Liao:2021yog,Bell:2021ihi,Araujo:2022wjh,Cox:2022ekg,Adams:2022zvg}. However, while the theoretical description of the Migdal effect has improved significantly in the past few years~\cite{Dolan:2017xbu,Bell:2019egg,Baxter:2019pnz,Essig:2019xkx,Liang:2019nnx,Liu:2020pat,Kahn:2020fef,Flambaum:2020xxo,Bell:2021zkr,Acevedo:2021kly,Wang:2021oha,Blanco:2022pkt,Cox:2022ekg,Liang:2020ryg,Knapen:2020aky,Liang:2022xbu}, a general description of the Migdal effect for low-mass DM ($m_\chi \lesssim 50$~MeV) in solid-state crystal (semiconductor) targets is still lacking.  In this region, the DM transfers very little energy and momentum to the nucleus and is unable to dislodge it from its lattice site.  A proper description of this region, which includes the possibility of phonon production, is required.  In this paper, we will provide such a description based on an effective field theory (EFT). 

The outline of the paper is as follows. In Sec.~\ref{sec:overview}, we review the kinematics of the Migdal effect in semiconductors, provide an overview of the various theoretical calculations available in the literature, and contrast these with our effective field theory approach.   In Sec.~\ref{sec:fermi_theory}, we derive the effective Hamiltonian for the Migdal effect in semiconductors, and use it to derive an expression for the DM scattering rate in terms of the dynamic structure factor, $S(\bm q,E)$, a measurable quantity that depends only on the material's properties. In Sec.~\ref{sec:dynamic_structure_function}, we discuss how to obtain the dynamic structure factor from data and theoretical estimates. In Sec.~\ref{sec:results}, we show for low DM masses the total DM--nucleus Migdal scattering rates, as well as the differential rates, in electron recoil energy and in phonon energies for silicon and germanium targets. In Sec.~\ref{sec:conclusions}, we summarize our main findings and conclude with future directions. 
In the Appendices, we present additional  derivations, as well as a discussion on the prospects of using neutron scattering to obtain comprehensive data quantifying the dynamic structure factor.

\section{The Migdal effect in semiconductors in the low momentum regime}\label{sec:overview}

Semiconductors are excellent DM direct-detection targets.  
Their bandgaps, denoted here by $\omega_{\rm g}$, are $\mathcal{O}(\text{eV})$, so that DM with masses as low as $\mathcal{O}$(MeV) (which have kinetic energies $\frac{1}{2} m_\chi v_\chi^2 \sim  \mathcal{O}(\text{eV})$, assuming $v_\chi \sim 10^{-3}$) can deposit sufficient energy to excite an electron across the bandgap and create an electron--hole pair.  
The excitation energies are smaller than the ionization threshold of free atoms, which are $\mathcal{O}(10 \text{ eV})$. 
However, extending the description of the Migdal effect for a free atom  
to semiconductor targets requires a careful treatment of the fact that the nuclei are bound~\cite{Knapen:2020aky, Liang:2022xbu}, 
the electrons are not localized~\cite{Liang:2019nnx}, and the boost operator inducing the ionization in the free-atom treatment cannot simply be applied in a crystal with a preferred rest frame~\cite{Essig:2019xkx}. 
Recent work~\cite{Knapen:2020aky, Liang:2020ryg} has shown that the Migdal effect in semiconductors can be described by a process of DM--nucleus scattering in association with a nucleus--electron Coulomb interaction, the latter being responsible for the ionization of one of the electrons in the material. In particular, it has been shown that the well-studied energy loss function (ELF) can be used to encode the screening from valence electrons.

The recent work~\cite{Knapen:2020aky, Liang:2020ryg} resolves two of the three subtleties introduced by semiconductors, but does not fully address the bound nature of the nuclei. In the treatment presented in~\cite{Knapen:2020aky, Liang:2020ryg}, the final state of the nucleus is treated as a plane wave, hence ignoring the effects coming from the rest of the crystal lattice and any collective effects (phonons) that can be generated along with the Migdal electron during the DM scattering event. The approximation of treating the final state of the nucleus as a plane wave is usually dubbed \textit{impulse approximation} (ia), although within this approximation, the treatment of the initial state of the nucleus in the literature varies. In what \cite{Knapen:2020aky} call impulse approximation, the initial state of the nucleus is treated as the ground state of an harmonic crystal. In \cite{Liang:2020ryg}, instead, impulse approximation refers to treating both the initial and final states as plane waves. In this paper, our definition of the impulse approximation is consistent with that in~\cite{Knapen:2020aky}, whereas we label the treatment in~\cite{Liang:2020ryg} as the \textit{free-ion approximation}. 
To summarize, in this paper, we assume:

\vskip 2mm
\begin{tabular}{r l}
     impulse approx.: & final (initial) state of nucleus is plane wave (harmonic crystal) \\
     free-ion approx.: &  final (initial) state of nucleus is plane wave (plane wave)
\end{tabular}
\vskip 2mm

\begin{figure}
    \centering
    \includegraphics[width=0.85\textwidth]{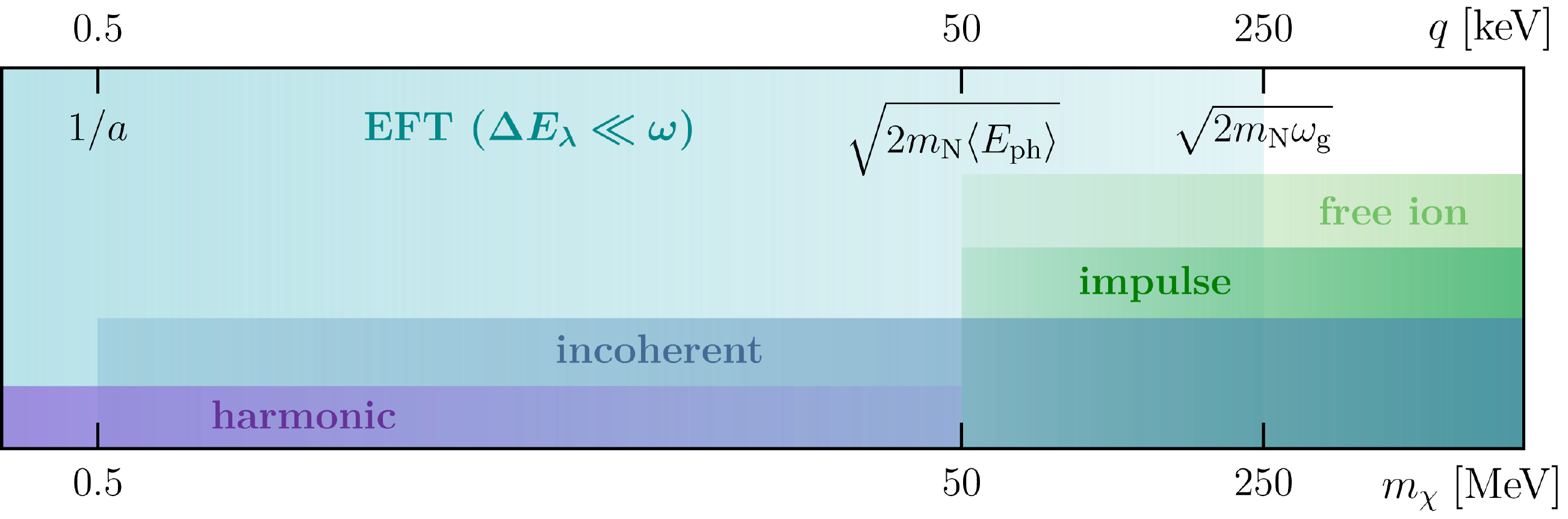}
    \caption{Schematic representation of the regimes of validity of the different approximations discussed in this work. Here $m_{\rm N}$ is the nucleus mass, $a$ the lattice spacing, $\langle E_{\rm ph} \rangle$ the typical average phonon energy, and $\omega_{\rm g}$ the semiconductor gap. 
    Both the impulse and the free-ion approximation are valid when the momentum transfer is much larger than the typical momentum scale set by the crystal, $q\gg \sqrt{2m_{\rm N} \langle E_{\rm ph} \rangle}$. However, the free-ion approximation is parametrically slower than the impulse approximation at converging to the corresponding approximate value. Corrections to the free-ion approximations are, in fact, proportional to $\sqrt{\langle E_{\rm ph} \rangle / E_r}$, while corrections to the impulse approximation are proportional to $\langle E_{\rm ph}\rangle/E_r$ (see Sec.~\ref{sec:freeion} for details).
    The harmonic approximation is, instead, valid in the opposite regime, $q \ll \sqrt{2m_{\rm N} \langle E_{\rm ph} \rangle}$, where corrections to the atomic potential that are higher than quadratic in the positions are neglected, and the displacement of the nucleus by the DM scatter is small compared to the inter-atomic spacing. The incoherent approximation applies to processes where the momentum exchanged is, instead, much larger than the inverse lattice spacing, $q\gg 1/a$, when the scattering particle is able to discriminate each single nucleus and hence does not scatter coherently off multiple lattice sites. Finally, the EFT presented in this work is valid when the separation between the lattice energy levels excited by the DM, $\Delta E_\lambda$, is much smaller than the energy of the Migdal electron, $\omega$---i.e., for momentum transfers $q\ll \sqrt{2m_{\rm N} \omega_{\rm g}}$.
    Note that close to the boundaries of the indicated regions of validity, the approximations will receive $\mathcal{O}(1)$ corrections. 
    The numerical values are estimated for the case of silicon, for which $m_{\rm N} \simeq 28$~GeV, $\omega_{\rm g} \simeq 1.2$~eV, and $\langle E_{\rm ph} \rangle \simeq 0.04$~eV. The typical DM velocity is taken to be $v_\chi \simeq 10^{-3}$.}
    \label{fig:regimes}
\end{figure}

Both approximations are only good approximations when DM--nucleus collision happens over time scales, $t_{\rm coll}$, much shorter than the typical time scale characterizing the dynamics of the lattice, i.e., for
\begin{align} \label{eq:impulse-timescale}
    t_{\rm coll} \sim 1/E_r \ll t_{\rm ph} \sim 1/\langle E_{\rm ph} \rangle\, \qquad \textrm{(impulse approximation)\,,}
\end{align} 
with the free-ion approximation breaking down faster than the impulse approximation (see Fig.~\ref{fig:regimes}). 
Here $\langle E_{\rm ph} \rangle$ is the average phonon energy and $E_r$ is the energy of the recoiling nucleus,
\begin{subequations}
    \begin{align}
        E_r ={} & \frac{q^2}{2m_{\rm N}}  \label{eq:Er} \\
        \sim{} & \frac{m^2_{\chi} v^2_{\chi}}{2m_{\rm N}} \simeq 0.04~\textrm{eV}~\left(\frac{m_\chi}{50~\textrm{MeV}}\right)^2 \left(\frac{28~\textrm{GeV}}{m_{\rm N}}\right) \,, \label{eq:Er2}
    \end{align}
\end{subequations}
where $q$ is the momentum lost by the DM, $m_{\rm N}$ is the mass of the nucleus (taken to be the approximate mass of a silicon nucleus in the second line), and we assume a DM velocity of $v_\chi\sim 10^{-3}$.  
Combining Eqs.~(\ref{eq:impulse-timescale}) and (\ref{eq:Er}), the impulse approximation is valid for $q \gg \sqrt{2 m_N \langle{E_\textrm{ph}}\rangle}$.  
For silicon and germanium, the average phonon energies are $\langle E^{\text{Si}}_{\text{ph}} \rangle \simeq 0.04 \text{ eV}$ and $\langle E^{\text{Ge}}_{\text{ph}} \rangle \simeq 0.025 \text{ eV}$, respectively.  
Combining Eqs.~(\ref{eq:impulse-timescale}) and (\ref{eq:Er2}), we find
\begin{align}\label{eq:impulse-timescale2}
m_\chi \gg 50~\textrm{MeV} \sqrt{\frac{m_{\rm N}}{28~\textrm{GeV}}}
\sqrt{\frac{\langle E_{\text{ph}} \rangle}{0.04 \, \text{eV}}}  \qquad \textrm{(impulse approximation)\,.} 
\end{align}
We see that the condition in Eq.~(\ref{eq:impulse-timescale2}) is no longer satisfied for DM candidates with masses $m_\chi \lesssim 50$~MeV. 
Given that DM candidates with masses as small as $\mathcal{O}$(MeV) are still kinematically able to excite an electron in a semiconductor via the Migdal effect, it is imperative to have a robust theoretical understanding of the Migdal effect also for $1~\textrm{MeV}\lesssim m_\chi \lesssim 50~\textrm{MeV}$, where the impulse approximation fails. 

In Fig.~\ref{fig:regimes}, we show a schematic representation of the break-down of the impulse approximation as a function of the momentum transfer and of the dark matter mass, and  overlay it with the regime of validity of the EFT approach we employ in this work to extend a description of the Migdal effect in semiconductors to the low-mass regime. Fig.~\ref{fig:regimes} also shows the regime of validity of the incoherent as well as the harmonic approximations, introduced in Sec.~\ref{sec:dynamic_structure_function}, and which are relevant for the numerical results presented in Sec.~\ref{sec:results}. 

In this work, we reformulate the low-momentum Migdal effect in semiconductors in terms of an effective field theory, which is insensitive to the complicated short-distance details of the crystal lattice. 
We are able to do so by exploiting the separation of scales between the Migdal ionization energy, $\omega \gtrsim  \mathcal{O}(1 \text{ eV})$, and the typical energy of the lattice excitations, $E_{\rm ph} \lesssim \mathcal{O}(0.1 \text{ eV)}$. By integrating out the short-distance (high-momentum) vibrational modes, we simplify the description of the Migdal effect in semiconductors from a second-order effect in old fashioned perturbation theory (a DM--nucleus interaction plus a nucleus--electron Coulomb interaction) to a first order effect with an effective Hamiltonian, the latter being independent of the complicated interactions between the ions of the crystal. 

The low-energy effective theory allows for a description of the final rate that only depends on the dynamic structure factor of the target material, $S(\vec{q}, E)$---a measurable quantity that is fully characterized depends by the momentum transferred to the lattice, $\bm q$, and the energy deposited to the lattice, $E$. We make no assumptions about the initial, intermediate, or final state of the nucleus in this EFT approach. We accurately quantify the double differential 
Migdal emission rate $dR/(d \omega dE)$
in a semiconductor, where $\omega$ is the energy of the ionized electron, 
an observable relevant for next generation detectors with sensitivity to single- and multi-phonon excitations. 
Interestingly, we use sum rules to show that the single differential rate, $dR/d\omega$, is independent of the details of the lattice dynamics, and equivalent to the same result obtained in the free-ion approximation.

This extends the range of validity of the description of the Migdal effect in semiconductors to all kinematically accessible DM masses, $m_\chi \gtrsim 1 \text{ MeV}$. Additionally, the use of an effective field theory allows for the systematic inclusion of higher order terms in the small frequency expansion. 

While our work was in progress, a proposal for a possible description on how to take into effect the bound nucleus nature for small DM masses appeared in~\cite{Liang:2022xbu}. Inspired by standard field theory, the authors derive Feynman rules for multi-phonon and electron--phonon interactions, and show that, when the number of phonons in the final state is large, their results are well reproduced by the impulse approximation. This approach, however, is only applicable within the harmonic approximation for the lattice dynamics, which treats the inter-atomic potential as purely quadratic in the positions, and where the theory can be solved exactly. Our derivation, instead, holds generally for any crystal. 
As we will show, when evaluated within the impulse approximation, our result reduces to that of~\cite{Knapen:2020aky}, while within the harmonic approximation,
our analytical result agrees with~\cite{Liang:2022xbu}.\footnote{The numerical results shown in v1 and v2 of \cite{Liang:2022xbu} differ from our results in the low mass regime $(1  \text{ MeV} - 50 \text{ MeV})$. The authors omitted a term in their numerical calculation, which they included in v3 after our correspondence. We discuss the relevant term in more detail in Section~\ref{sec:results}.}

\section{An effective theory for the Migdal effect}
\label{sec:fermi_theory}

\subsection{The effective Hamiltonian}

As highlighted in~\cite{Knapen:2020aky}, the Migdal emission of an electron in a solid state material formally involves two interactions, one between the DM particle and the nuclei in the crystal, and another between the nuclei and the electrons in the system, which induces their ionization. These two interactions are mediated by a lattice vibrational degree of freedom. For a realistic material, the intermediate lattice mode inducing the ionization is a complicated eigenstate of the full lattice Hamiltonian, whose complete description is prohibitive. For this reason, it is often convenient to simplify the treatment by making some assumptions, most notably the impulse~\cite{Knapen:2020aky} and/or the harmonic~\cite{Liang:2022xbu} approximations. While they do make the problem tractable, both these approximations break down at given energy scales (Fig.~\ref{fig:regimes}) and have corrections that are difficult to quantify.

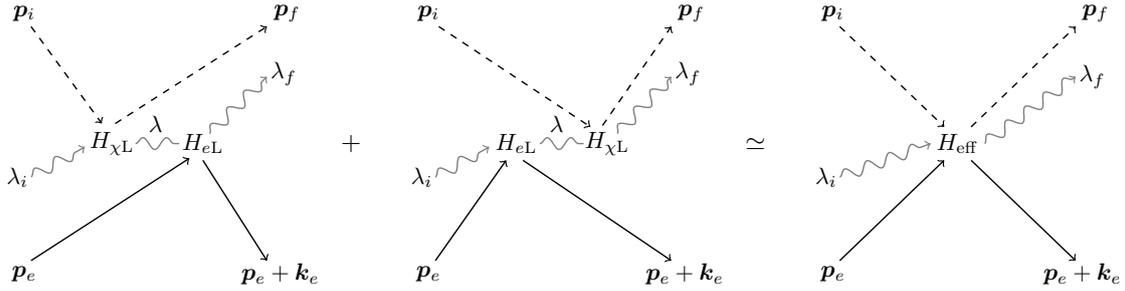
\begin{figure}
    \centering
    \resizebox{1\textwidth}{!}{
        \begin{tikzpicture}
            \draw[->,semithick] (-1,0.0) -- (1.35,1.53);
            \draw[->,gray,snake it,semithick] (-1,1.25) -- (-0.2,1.7);
            \draw[->,dashed,semithick] (-1,3.5) -- (0.05,2.05);
            \draw[gray,snake it,semithick] (0.55,1.75) -- (1.2,1.75);
            \draw[->,gray,snake it,semithick] (1.65,1.9) -- (2.5,2.75);
            \draw[->,dashed,semithick] (0.25,2.05) -- (2.5,3.5);
            \draw[->,semithick] (1.55,1.5) -- (2.5,0.0);
            
            \node at (-1.1,-0.2) {\small $\bm{p}_e$};
            \node at (-1.2,1.25) {\small $\lambda_i$};
            \node at (-1.1,3.7) {\small $\bm{p}_i$};
            \node at (2.7,-0.2) {\small $\bm{p}_e + \bm{k}_e$};
            \node at (2.75,2.8) {\small $\lambda_f$};
            \node at (2.8,3.7) {\small $\bm{p}_f$};
            \node at (0.85,2.05) {\small $\lambda$};
            
            \node at (0.2,1.75) {\small $H_{\chi \rm L}$};
            \node at (1.55,1.75) {\small $H_{e \rm L}$};
            
            \node at (3.75,1.75) {$+$};
            
            \draw[->,semithick] (5,0.0) -- (6.05,1.5);
            \draw[->,gray,snake it,semithick] (5,1.25) -- (5.8,1.7);
            \draw[gray,snake it,semithick] (6.55,1.75) -- (7.2,1.75);
            \draw[->,semithick] (6.3,1.5) -- (8.5,0.0);
            \draw[->,gray,snake it,semithick] (7.7,1.9) -- (8.5,2.75);
            \draw[->,dashed,semithick] (5.05,3.5) -- (7.3,2.0);
            \draw[->,dashed,semithick] (7.48,2.02) -- (8.5,3.5);
            
            \node at (4.9,-0.2) {\small $\bm{p}_e$};
            \node at (4.8,1.25) {\small $\lambda_i$};
            \node at (4.9,3.7) {\small $\bm{p}_i$};
            \node at (8.7,-0.2) {\small $\bm{p}_e + \bm{k}_e$};
            \node at (8.75,2.8) {\small $\lambda_f$};
            \node at (8.8,3.7) {\small $\bm{p}_f$};
            \node at (6.8,2.05) {\small $\lambda$};
            
            \node at (6.2,1.75) {\small $H_{e \rm L}$};
            \node at (7.55,1.75) {\small $H_{\chi \rm L}$};
            
            \node at (9.75,1.75) {$\simeq$};
            
            \draw[->,semithick] (11,0.0) -- (12.55,1.5);
            \draw[->,gray,snake it,semithick] (11,1.25) -- (12.35,1.75);
            \draw[->,dashed,semithick] (11,3.5) -- (12.55,2.0);
            \draw[->,semithick] (12.95,1.5) -- (14.5,0.0);
            \draw[->,gray,snake it,semithick] (13.15,1.75) -- (14.5,2.75);
            \draw[->,dashed,semithick] (12.95,2.0) -- (14.5,3.5);
            
            \node at (10.9,-0.2) {\small $\bm{p}_e$};
            \node at (10.8,1.25) {\small $\lambda_i$};
            \node at (10.9,3.7) {\small $\bm{p}_i$};
            \node at (14.6,-0.2) {\small $\bm{p}_e + \bm{k}_e$};
            \node at (14.75,2.75) {\small $\lambda_f$};
            \node at (14.8,3.7) {\small $\bm{p}_f$};  
            
             \node at (12.75,1.75) {\small $H_{\rm eff}$};
        \end{tikzpicture}
    }
    \caption{Schematic representation of the EFT procedure applied to the result of old-fashioned perturbation theory. Each line corresponds to a state in the Hilbert spaces of the dark matter (\textbf{dashed}), electron (\textbf{solid}), and crystal lattice (\textbf{wavy}). The intermediate lattice mode has high frequency and, when integrated out, it leads to an effective Hamiltonian that is local in time and independent on the complicated dynamics of the lattice.}
    \label{fig:Fermi}
\end{figure}

Luckily, in the context of interest to us ($m_{\chi} \lesssim \mathcal{O}(100 \text{ MeV})$), the problem is characterized by a separation of scales: the energy of the Migdal electron is at least a few eV,\footnote{The approximate thresholds to excite two electron--hole pairs in silicon and germanium are $5 \text{ eV}$ and $3.7 \text{ eV}$, respectively.} which is much larger than the typical energy of the lattice degrees of freedom excited in the process, which is of the order of tens to hundreds meV. As we will show, in this regime, the lattice degree of freedom mediating the ionization is highly off-shell. Following the standard EFT approach, one can integrate it out, and the final effective interaction is independent of the complicated microscopic dynamics of the crystal lattice. 
Operationally, this is achieved by expanding the matrix element for large electron energy, $\omega$, reducing the second order interaction to a first order interaction. As we show below, the latter is such that it does not depend on the inter-atomic potential of the crystal lattice.
The schematic EFT procedure is represented in Fig.~\ref{fig:Fermi}.
 
Let us start by treating the problem in full generality. We will mostly adopt the notation of~\cite{Knapen:2020aky}. In old-fashioned perturbation theory, the Migdal emission is a second order process, whose matrix element is given by,
\begin{align} \label{eq:M}
    \begin{split}
        \mathcal{M}_{fi} ={}& \sum_{\lambda} \bigg[ \frac{\langle \lambda_f,\bm{p}_e + \bm{k}_e | H_{e\rm L} | \lambda, \bm{p}_e \rangle \langle \bm{p}_f, \lambda | H_{\chi\rm L}|\bm{p}_i,\lambda_i \rangle}{\omega + E_{\lambda_f} - E_\lambda} \\
        & \quad\, + \frac{\langle \bm{p}_f, \lambda_f |H_{\chi\rm L}|\bm{p}_i,\lambda \rangle \langle \lambda, \bm{p}_e + \bm{k}_e | H_{e\rm L}|\lambda_i, \bm{p}_e \rangle}{E_{\lambda_i} - E_\lambda - \omega} \bigg] \,.
    \end{split}
\end{align}
Here, $\bm{p}_i$ and $\bm{p}_f$ are the initial and final DM momenta, $\bm{p}_e$ and $\bm{p}_e + \bm{k}_e$ the initial and final electron (crystal) momenta, and $\lambda_i$, $\lambda_f$, and $\lambda$ are the initial, final, and intermediate lattice states, respectively. Moreover, $\omega$ is the energy gained by the electron. The Hamiltonian $H_{e \rm{L}}$ represents the coupling between an electron and the lattice, while $H_{\chi \rm{L}}$ analogously represents the coupling of the DM to the lattice.

For the range of DM masses of interest to us, the electron energy is always substantially larger than the typical spacing between the lattice eigenvalues entering in Eq.~\eqref{eq:M}, $\omega \gg |E_{\lambda_f} - E_\lambda|\,, | E_{\lambda_i} - E_\lambda |$. This is because, for a given external kinematics, the matrix elements in the numerator will suppress the contribution from intermediate states that are too far away from $E_{\lambda_i}$ and $E_{\lambda_f}$.\footnote{This approximation clearly breaks down for sufficiently heavy DM particles, which are able to excite lattice modes whose energy is much higher than the electron's recoil energy---even kick a nucleus out of the crystal, in the extreme case. If we estimate the energy given to the crystal as the recoil energy of a free nucleus, $q^2/2m_{\rm N}$, the EFT breaks down when $q\sim \sqrt{2m_{\rm N} \omega_{\rm g}}$, with $\omega_{\rm g}$ the semiconductor bandgap. In the case of silicon and germanium, one has $\omega_{\rm g}\simeq 1.2$~eV  and $\omega_{\rm g} \simeq 0.7$~eV, respectively. This corresponds to $m_\chi \gtrsim 250$~MeV and $m_\chi \gtrsim 320$~MeV. We note that, in principle, our EFT is always valid for sufficiently energetic electrons, $\omega \gg q^2/(2m_{\rm N})$. Here we take the most conservative value, $\omega \sim \omega_{\rm g}$.} We can then expand the matrix element, which reads
\begin{align}
     \begin{split}
         \mathcal{M}_{fi} ={}& \sum_\lambda \Bigg[ \bigg( \frac{1}{\omega} - \frac{E_{\lambda_f} - E_\lambda}{\omega^2} \bigg) \langle \lambda_f,\bm{p}_e + \bm{k}_e | H_{e\rm L} | \lambda, \bm{p}_e \rangle \langle \bm{p}_f, \lambda | H_{\chi\rm L}|\bm{p}_i,\lambda_i \rangle \\
         & - \bigg( \frac{1}{\omega} - \frac{E_{\lambda} - E_{\lambda_i}}{\omega^2} \bigg) \langle \bm{p}_f, \lambda_f |H_{\chi\rm L}|\bm{p}_i,\lambda \rangle \langle \lambda, \bm{p}_e + \bm{k}_e | H_{e\rm L}|\lambda_i, \bm{p}_e \rangle \Bigg] + \mathcal{O}\big( 1/\omega^3 \big) \, ,
          \\
          ={}& \frac{1}{\omega^2} \bigg[ \langle \bm{p}_f, \lambda_f, \bm{p}_e + \bm{k}_e | \big( H_{e\rm L} H_{\rm L} - H_{\rm L} H_{e \rm L} \big) H_{\chi \rm L} - H_{\chi \rm L} \big( H_{e \rm L} H_{\rm L} - H_{\rm L} H_{e \rm L} \big) | \bm{p}_i, \lambda_i, \bm{p}_e \rangle \bigg]  \\
          & +\mathcal{O}\big(1 / \omega^3 \big) \,,
     \end{split}
\end{align}
where when going to the second equation we used the fact that the $|\lambda \rangle$'s are eigenstates of the unperturbed lattice Hamiltonian, $H_{\rm L}$, as well as the associated completeness relation, $\sum_\lambda |\lambda \rangle \langle \lambda | = \mathbbm{1}$. One deduces the effective interaction Hamiltonian, 
\begin{align} \label{eq:Heff}
    H_{\rm eff} = \frac{1}{\omega^2} \big[ H_{\chi \rm L}, [ H_{\rm L}, H_{e \rm L} ] \big] + \mathcal{O} \big( 1/\omega^3 \big) \,,
\end{align}
such that $\mathcal{M}_{fi} = \langle \lambda_f, \bm p_e + \bm k_e, \bm p_f| H_{\rm eff} | \lambda_i, \bm p_e, \bm p_i \rangle + \mathcal{O}\big( 1/ \omega^3 \big)$.
We can now specialize our result to the case at hand. The DM--lattice and electron--lattice interactions are described by~\cite{Knapen:2020aky},
\begin{subequations}
    \begin{align}
        H_{\chi \rm L} ={}& - \sum_I \frac{g_\chi g_{\rm N}}{4\pi} \frac{e^{-m_\phi | \vec{x}_\chi - \vec{x}_I|}}{|\vec{x}_\chi - \vec{x}_I|} \equiv \sum_I H_{\chi \rm L}^{(I)} \,, \\
        H_{e \rm L} ={}& - \frac{4\pi \alpha}{V} \sum_{I} \sum_{\vec{K},{\vec{K}}^\prime} \sum_{\vec{k}}  \frac{\epsilon_{\vec{K}\vec{K}^\prime}^{-1}(\vec{k},\omega)Z(|\bm{k}+\bm{K}^\prime|)}{|\vec{k}+\vec{K}||\vec{k}+\vec{K}^\prime|} e^{i(\vec{k}+\vec{K})\cdot \vec{x}_e
        } e^{-i(\vec{k}+\vec{K}^\prime)\cdot\vec{x}_I} \equiv \sum_{I} H_{e \rm L}^{(I)} \,, \label{eq:HeL}
    \end{align}
\end{subequations}
where $\alpha$ is the fine structure constant, $V$ the volume of the material, and $Z(k)$ the effective atomic number accounting for the tightly bound core electrons. Its momentum dependence encodes the fact that, at high momentum, one probes deeper into the core electrons, effectively experiencing a larger positive charge. Moreover, $\epsilon_{\vec{K}\vec{K}^\prime}^{-1}(\vec{k},\omega)$ is the longitudinal dielectric matrix (symmetrized over $\vec{K}$ and $\vec{K}^\prime$), $\vec{K}^{(\prime)}$ is a reciprocal lattice vector, while $\vec{k}$ is confined to the first Brillouin zone. Finally, $\vec{x}_\chi$, $\vec{x}_e$, and $\vec{x}_I$ are respectively the positions of the DM, the Migdal electron, and the $I$-th nucleus. For pedagogical reasons, we report the derivation of the electron--lattice Hamiltonian in Appendix~\ref{app:HeL}. For the DM--lattice interaction, we are assuming for concreteness a DM--nucleus interaction due to a Yukawa coupling with a scalar (vector) mediator of mass $m_\phi$ ($m_A$), hence allowing for both short and long range forces. Specifically, we consider an interaction Lagrangian given by $\mathcal{L^{\phi}}_{\rm int} = g_\chi \phi \bar \chi \chi + g_{\rm N} \phi \bar N N$ ($\mathcal{L^{A}}_{\rm int} = g_\chi A_{\mu} \bar \chi \gamma^{\mu} \chi + g_{\rm N} A_{\mu} \bar N \gamma^{\mu} N$,), where $N$ denotes the nucleus.

The lattice Hamiltonian can generically be written as 
\begin{align}
    H_{\rm L} = - \sum_I \frac{\nabla_I^2}{2m_{\rm N}} + U\big( \{ \vec{x}_I \} \big) \,,
\end{align}
where $\bm{\nabla}_I$ is the gradient with respect to the $\vec{x}_I$ position and $m_{\rm N}$ the nucleus mass. From now on, we assume identical atoms to keep the notation simple though our formalism can easily be generalized to multi-atomic crystals such as GaAs~\cite{Derenzo:2016fse,Derenzo:2018plr}. Importantly, for a realistic material the inter-atomic potential, $U$, is in general a complicated anharmonic function. Nonetheless, the effective Hamiltonian~\eqref{eq:Heff} only depends on commutators. Since $H_{e \rm L}$ and $H_{\chi \rm L}$ are $c\,$-numbers, they commute with the potential, and the effective interaction is therefore independent of the potential, 
\begin{align}
    H_{\rm eff} = \frac{1}{m_{\rm N} \omega^2} \sum_{I} \bm{\nabla}_I H_{\chi \rm L}^{(I)} \cdot \bm{\nabla}_I H_{e \rm L}^{(I)} + \mathcal{O} \big(1/\omega^3\big) \,.
\end{align}

\subsection{The rate for Migdal emission}

It is now straightforward to show that, given the above Hamiltonian, the corresponding matrix element is,
\begin{align*}
    \begin{split}
        \mathcal{M}_{fi} ={}& \! - \frac{4 \pi \alpha g_\chi g_{\rm N}}{m_{\rm N} \omega^2 V^2} \sum_{I} \! \sum_{\vec{K}, \vec{K}^\prime} \! \sum_{\vec{k}} \frac{\vec{q}\!\cdot\!(\vec{k} + \vec{K}^\prime )}{q^2+m_\phi^2} \frac{\epsilon^{-1}_{\vec{K}\vec{K}^\prime} (\vec{k},\omega)}{|\vec{k}+\vec{K}||\vec{k}+\vec{K}^\prime|} Z(|\bm{k}+\bm{K}^\prime|) \\
        & \times \langle \lambda_f | e^{i(\vec{k} + \vec{K}^\prime - \vec{q})\cdot\vec{x}_I} | \lambda_i \rangle \langle \vec{p}_e + \vec{k}_e | e^{i(\vec{k} + \vec{K})\cdot\vec{x}_e} | \vec{p}_e \rangle \\
        ={}& \! - \frac{4 \pi \alpha g_\chi g_{\rm N}}{m_{\rm N} \omega^2 V^2} \sum_{I} \! \sum_{\vec{K}, \vec{K}^\prime} \!\! \frac{\vec{q}\!\cdot\!(\vec{k}_e + \vec{K}^\prime )}{q^2+m_\phi^2} \frac{\epsilon^{-1}_{\vec{K}\vec{K}^\prime} (\vec{k}_e,\omega) }{|\vec{k}_e + \vec{K}||\vec{k}_e + \vec{K}^\prime|} Z(|\bm{k}_e+\bm{K}^\prime|) \\
        & \times \langle \lambda_f | e^{i(\vec{k}_e + \vec{K}^\prime - \vec{q})\cdot\vec{x}_I} | \lambda_i \rangle [ \vec{p}_e + \vec{k}_e | e^{i(\vec{k}_e + \vec{K})\cdot\vec{x}_e} | \vec{p}_e ]_\Omega \,, 
    \end{split}
\end{align*}
where $\vec{q} \equiv \vec{p}_i - \vec{p}_f$ is the momentum released by the DM, and in the second line we have used Bloch's theorem for the electron state~\cite{Knapen:2020aky}. In particular, $|\vec{p}_e]_\Omega$ represents the (periodic) Bloch wave function~\citep[e.g.,][]{ashcroft2020solid}, and the matrix element between two electronic states is to be computed over the volume of a primitive cell, $\Omega$, as indicated by the subscript.

At zero temperature, $|\lambda_i\rangle$ is simply the ground state of the lattice Hamiltonian, while the initial electron state, $|\bm{p}_e]_\Omega$, follows a distribution $f_b(\bm{p}_e)$, with $b$ labeling the electronic branch.
Fermi's golden rule then returns the following rate,
\begin{subequations}
    \begin{align} 
            \frac{d\Gamma}{d\omega} ={}& \frac{4\alpha}{V^3} \bigg( \frac{ g_\chi g_{\rm N}}{m_{\rm N} \omega^2} \bigg)^{\!\!2} \sum_{\vec{q}} \sum_{\vec{k}_e} \sum_{\vec{K},\vec{K}^\prime} \sum_{\vec{Q},\vec{Q}^\prime} \frac{\vec{q}\!\cdot\!(\vec{k}_e + \vec{K}^\prime) \, \vec{q}\!\cdot\!(\vec{k}_e + \vec{Q}^\prime)}{\big( q^2 + m_\phi^2 \big)^2} \frac{Z(|\bm{k}_e+\bm{K}^\prime|) Z(|\bm{k}_e+\bm{Q}^\prime|)}{|\vec{k}_e + \vec{K}^\prime||\vec{k}_e + \vec{Q}^\prime|}  \notag  \\
            & \times \frac{4\pi^2\alpha}{V} \sum_{b,b^\prime} f_b(\bm{p}_e) \frac{ [\vec{p}_e| e^{-i \vec{Q}\cdot \vec{x}_e} | \vec{p}_e + \vec{k}_e ]_\Omega  \, [\vec{p}_e +\vec{k}_e| e^{i\vec{K}\cdot\vec{x}_e} | \vec{p}_e ]_\Omega }{|\vec{k}_e + \vec{K}||\vec{k}_e + \vec{Q}|} \notag \\ 
            & \times \delta(\omega - E_{b^\prime,\vec{p}_e+\vec{k}_e} - E_{b,\vec{p}_e} ) \epsilon_{\vec{K}\vec{K}^\prime}^{-1}(\vec{k}_e,\omega)\epsilon_{\vec{Q}\vec{Q}^\prime}^{-1 *}(\vec{k}_e,\omega) \label{eq:dGammaA}
            \\[1em]
            & \times \sum_{\lambda_f} \, \langle \lambda_i | \sum_I e^{-i(\vec{k}_e + \vec{Q}^\prime - \vec{q})\cdot\vec{x}_I} | \lambda_f \rangle \langle \lambda_f | \sum_J e^{i(\vec{k}_e + \vec{K}^\prime - \vec{q})\cdot\vec{x}_J} | \lambda_i \rangle \notag \\
            & \times (2\pi)\delta \left( E_{\vec{p}_i} - E_{\vec{p}_f} + E_{\lambda_i} - E_{\lambda_f} - \omega \right) \notag \\
        \begin{split} \label{eq:dGammaB}
            ={}& \frac{4\alpha}{V^3} \bigg( \frac{g_\chi g_{\rm N}}{m_{\rm N} \omega^2} \bigg)^{\!\!2} \sum_{\vec{q}} \sum_{\vec{k}_e} \sum_{\vec{K},\vec{Q}} \frac{\vec{q}\!\cdot\!(\vec{k}_e + \vec{K}) \, \vec{q}\!\cdot\!(\vec{k}_e + \vec{Q})}{\big( q^2 + m_\phi^2 \big)^2} \frac{ Z(|\bm{k}_e+\bm{K}|) Z(|\bm{k}_e+\bm{Q}|) }{|\vec{k}_e + \vec{K}||\vec{k}_e + \vec{Q}|}  \\
            & \times \text{Im}\big(\!-\epsilon_{\vec{K}\vec{Q}}^{-1}(\vec{k}_e,\omega)\big) \sum_{\lambda_f} \, \langle \lambda_i | \sum_I e^{-i(\vec{k}_e + \vec{Q} - \vec{q})\cdot\vec{x}_I} | \lambda_f \rangle  \\
            & \times \langle \lambda_f | \sum_J e^{i(\vec{k}_e + \vec{K} - \vec{q})\cdot\vec{x}_J} | \lambda_i \rangle (2\pi) \delta \left( E_{\vec{p}_i} - E_{\vec{p}_f} + E_{\lambda_i} - E_{\lambda_f} - \omega \right) \,, 
        \end{split} 
    \end{align}
\end{subequations}
where $E_{b,\bm{p}_e}$ and $E_{b^\prime,\bm{p}_e+\bm{k}_e}$ are the initial and final electron energies, and $E_{\vec{p}_i}$ and $E_{\vec{p}_f}$ the initial and final DM energies. In the second equality we used Lindhard's formula for the imaginary part of the dielectric matrix~\cite{lindhard1954properties,adler1962quantum,Knapen:2020aky}, corresponding precisely to the second and third lines of Eq.~\eqref{eq:dGammaA}.\footnote{The generalization of Lindhard's formula to the reciprocal lattice actually includes an extra term, associated to the possibility of absorbing an initially free electron (cf. Appendix~A in~\cite{Knapen:2020aky}). In the ground state the density of initially free electrons is negligible, and this term does not contribute.} The function $\text{Im}\big( \! - \epsilon_{\bm K \bm Q}^{-1}(\bm k_e,\omega) \big)$ is the ELF.

Now, in the space of reciprocal lattice vectors, the ELF has small off-diagonal elements~\cite{Knapen:2020aky}. For this reason, we can set $\bm{Q} \simeq \bm{K}$ in the last two lines of Eq.~\eqref{eq:dGammaB}.\footnote{For DM particle with $m_\chi \gtrsim 10$~MeV, this approximation becomes even better, since one has $q \gg |\bm{k}_e + \bm{K}|, |\bm{k}_e + \bm{Q}|$ and one can neglect the electron momenta in the last lines of Eq.~\eqref{eq:dGammaB}. The electron momentum supported by the ELF is, in fact, never larger than around $20$~keV~\cite{Knapen:2021bwg} for valence electrons which dominate the ionization probability for $m_\chi \lesssim 100$~MeV. } 
After doing that, we obtain our final expression for the rate of Migdal emission in a semiconductor:
\begin{align} \label{eq:dGammafinal}
\frac{d \Gamma}{d\omega} ={}& \frac{8 \pi \alpha N_{\rm T}}{V^3} \bigg( \frac{ g_\chi g_{\rm N}}{m_{\rm N} \omega^2} \bigg)^{\!\!2} \sum_{\vec{q}} \sum_{\vec{k}_e} \sum_{\vec{K},\vec{Q}} \frac{\vec{q}\!\cdot\!(\vec{k}_e + \vec{K}) \, \vec{q}\!\cdot\!(\vec{k}_e + \vec{Q})}{\big( q^2 + m_\phi^2 \big)^2} \frac{Z(|\bm{k}_e+\bm{K}|) Z(|\bm{k}_e+\bm{Q}|)}{|\vec{k}_e + \vec{K}||\vec{k}_e + \vec{Q}|} \notag \\
    & \times \text{Im}\big(\!-\epsilon_{\vec{K}\vec{Q}}^{-1}(\vec{k}_e,\omega)\big) S(\vec{q} - \bm{k}_e - \bm{K}, E_{\vec{p}_i} - E_{\vec{p}_f} - \omega) \,,
\end{align}
where $N_{\rm T}$ denotes the number of ions in the target, and $S(\vec{q},E)$ is the partial dynamic structure factor of the lattice~\citep[e.g.,][]{squires1996introduction}, i.e.
\begin{align} \label{eq:Sdef}
    S(\vec{q},E) \equiv \frac{1}{N_{\rm T}} \sum _{\lambda_f} \bigg| \langle \lambda_f | \sum_I    e^{i\vec{q} \cdot \vec{x}_I} | \lambda_i \rangle \bigg|^2  \delta\big(E_{\lambda_i} -E_{\lambda_f} -E\big) \,.
\end{align}
The advantage of expressing the rate in terms of the dynamic structure factor lies in the fact that, for many-body systems, it encodes the complicated inter-particle correlations, and only depends on the active degrees of freedom in the target system that are excited given a momentum transfer $\vec{q}$ \cite{Trickle:2019nya}.
In principle, one could define and attempt to measure a dynamic structure factor that incorporates the Migdal effect directly, by considering electronic degrees of freedom in its definition. This is practically challenging, both theoretically and experimentally. For this reason, in our discussion we keep the lattice and electronic degrees of freedom separated. The former are encoded in the structure factor, while the latter in the ELF.

\section{The dynamic structure factor}
\label{sec:dynamic_structure_function}

In principle, the dynamic structure factor can be measured through neutron scattering experiments, in which case no further assumptions would be necessary---Eq.~\eqref{eq:dGammafinal} can be used without detailed knowledge of the lattice dynamics. In practice, however, comprehensive data over the relevant range in $\vec{q}$ ($10 \text{ keV} - 100 \text{ keV}$) does not yet exist.\footnote{For $q < 10 \text{ keV}$, the dynamic structure factor is dominated by elastic scattering and the phonon density of states, which has been measured through neutron scattering experiments as well as calculated in ab initio calculations for silicon and germanium.} 

In Appendix~\ref{app:neutron_scattering}, we discuss the experimental prospects of taking those measurements. In this section, instead, we present an overview of the different approximation schemes that one can employ to model the structure factor.

\subsection{The incoherent and harmonic approximations}
\label{subsec:harmonic}

In the absence of data, we get as much analytical understanding as possible by evaluating $S(\bm{q},E)$ in a simple context. Specifically, we use the \textit{incoherent} and the \textit{harmonic} approximations.
The former neglects interference effects coming from the scattering off different lattice points. This is a fair assumption as long as the momentum transfer is substantially larger than the (inverse) inter-particle separation, a condition that is satisfied for all DM masses considered here (see Figure~\ref{fig:regimes}).
In particular, the incoherent approximation implies
\begin{align}
    S(\vec{q},E) \simeq  \frac{1}{N_{\rm T}}  \sum _{\lambda_f} \sum_I \left|  \langle \lambda_f |     e^{i\vec{q} \cdot \vec{x}_I} | \lambda_i \rangle \right|^2  \delta\big(E_{\lambda_i} -E_{\lambda_f} -E\big) \,.
\end{align}

The harmonic approximation, on the other hand, neglects corrections to the atomic potential, $U(\{\bm{x}_I\})$, that are higher than quadratic in the positions, and it holds when the atomic displacements are small compared to the inter-atomic spacing. This is only true for the lightest DM particles (say, $m_\chi \lesssim 30$~MeV~\cite{Campbell-Deem:2022fqm}), while for DM particles with rates dominated by higher momentum transfers, anharmonicities can give large corrections~\cite{Campbell-Deem:2022fqm}. This latter assumption is hence employed only to gain some analytical understanding of the physics, and to estimate the event rates in a simplified setting.

Let us briefly review how to construct the dynamic structure factor within the approximations outlined above.
The position of each atom in the lattice can be decomposed as $\bm{x}_I = \bm{x}_I^0 + \bm{u}_I$, where $\bm{x}_I^0$ is the equilibrium location of the $I$-th atom, and $\vec{u}_I$ its relative displacement. When evaluated for an harmonic crystal, Eq.~\eqref{eq:Sdef} simplifies to~\citep[e.g.,][]{squires1996introduction}
\begin{align} \label{eq:Sharmonic}
    S^{\rm (harm)}(\bm{q}, E) = \frac{1}{N_{\rm T}} \sum_I \int_{-\infty}^{\infty} \frac{dt}{2\pi} e^{- i E t} e^{-2W(\bm{q})} e^{\langle \bm{q} \cdot \hat{\bm{u}}_I(0)\,\bm{q}\cdot \hat{\bm{u}}_I(t)\rangle} \,,
\end{align}
where $\hat{\bm{u}}_I(t)$ is the displacement operator in the Heisenberg picture, responsible for the creation or annihilation of a single phonon degree of freedom. Moreover, $W(\bm q) \equiv \frac{1}{2} \big\langle (\bm{q}\cdot\hat{\bm{u}}_0(0))^2\big\rangle$ is the Debye--Waller factor. The correlators appearing in Eq.~\eqref{eq:Sharmonic} can both be determined in terms of the phonon density of states, $g(E_{\text{ph}})$, normalized as $\int_0^{\infty} dE_{\text{ph}} \, g(E_{\text{ph}}) = 1$. In particular, within the assumptions of identical atoms and cubic symmetry, the unequal time correlator is independent of the lattice point and of the direction of the exchanged momentum, and it is given by~\cite{schober2014introduction, doi:10.1119/1.1987042},
\begin{align} \label{eq:correlator}
    \langle {\bm{q} \cdot \hat{\vec{u}}_I(0) \, {\bm q}\cdot \hat {\bm u}_I(t) \rangle}  =\frac{q^2}{2m_{\rm N}}\int_{-\infty}^\infty dE_{\text{ph}}'\frac{g(E_{\text{ph}}')}{E_{\text{ph}}'} \big(n(E_{\text{ph}}')+1\big)e^{i E_{\text{ph}}' t} \equiv \frac{q^2}{2 m_{\rm N}} f(t) \,, 
\end{align} 
where $n(E_{\text{ph}})$ is a Bose--Einstein distribution. At the low temperature environments necessary for DM direct detection $ n \simeq 0$.\footnote{Many experiments operate at cryogenic temperatures, where $n\simeq 0$ is an excellent approximation; even for SENSEI, which operates at a much higher temperature of $T = 130 \text{ K}$, one gets (for a typical phonon energy of $\langle E_{\rm ph} \rangle = 40$~meV) $n \simeq 0.03$, which is also negligible.} 
From the equation above it follows that the Debye--Waller factor is
\begin{align} \label{eq:DW}
    2W({\bm q}) = \frac{q^2}{2m_{\rm N}} \int_0^{\infty} dE_{\text{ph}}' \frac{g(E_{\text{ph}}')}{E_{\text{ph}}'} \big(2n(E_{\text{ph}}') +1\big) \equiv \frac{q^2}{2m_{\rm N}} \big\langle E_{\rm ph}^{-1} \big\rangle   \,,  
\end{align}
where, from now on, we indicate averages over phonon energies as $\langle E_{\rm ph}^{n} \rangle$, for some $n$.

\subsubsection{Exact recursive procedure}
\label{subsubsec:recursion}

To compute the structure function, one can expand the factor $e^{\langle {\bm q}\cdot \hat {\bm u}_I(0)\,{\bm q}\cdot \hat {\bm u}_I(t)\rangle}$ order by order,
i.e. \cite{doi:10.1119/1.1987042, schober2014introduction, Berghaus:2021wrp, Campbell-Deem:2022fqm},
\begin{align} \label{eq:phononexact}
    S^{\rm (harm)}({\bm q}, E) &=    \int_{-\infty}^{\infty} \frac{dt}{2\pi} e^{- i E t} e^{-2W({\bm q})} \sum_p \frac{1}{p!}   \left(\frac{q^2}{2m_{\rm N}} f(t) \right)^p \\ & =   e^{-2W(\vec{q})} \left( \delta(E) +\frac{q^2}{2m_{\rm N}}\left(\frac{g(E)}{E} \big(n(E) +1\big)\right) + \sum_{p \geq 2} \frac{1}{p!}\left(\frac{q^2}{2m_{\rm N}} \right)^p T_p(E) \right) \notag \,.
\end{align}
Each term in the sum above corresponds to the contribution due to a final state with definite number of phonons---specifically, $p$ of them. The density of states depends on the material properties of the semiconductor, and can be obtained either experimentally (through neutron scattering, for example \cite{PhysRevB.5.3151, PhysRevB.102.174311}) or from ab initio calculations~\cite{SiDOS}. 
The functions $T_p(E)$ are, instead, determined by the recursion relation \cite{doi:10.1119/1.1987042, schober2014introduction, Berghaus:2021wrp}
\begin{align} \label{eq:Tprecursion}
    T_p(E)  = \int_{-\infty}^{+\infty} dE' \, T_1(E- E')T_{p-1}(E') \,, 
\end{align}
where $T_1(E) = g(E) \big(n(E) +1\big)/E$. This is computationally expensive, especially for large phonon numbers.

\subsubsection{Free-ion approximation} \label{sec:freeion}
The free-ion approximation assumes that the incoming and outgoing state of the nucleus are well described by a plane wave, in analogy to the treatment of a free atom. Strictly speaking this is never accurate for a nucleus in a quadratic potential, yet for energy deposition of $E_r \gg \langle E_{\rm ph} \rangle$ the approximation performs well. 
In this approximation the dynamic structure function is simply given by: 
\begin{align}\label{eq:free_ion}
    S^{\rm (\text{free ion})}({\bm q}, E) \simeq  \delta(E-E_r) \,.
\end{align}    
Note that both the impulse and the free-ion approximations are valid for $E_r \gg \langle E_{\rm ph} \rangle$. However, as we will show in Eqs.~\eqref{eq:expansion} and Eq.~\eqref{eq:impulse}, the corrections to the former scale by $\langle E_{\rm ph} \rangle t \sim \langle E_{\rm ph} \rangle/E_r$, while the latter is rather controlled by $\Delta/E_r = \sqrt{\langle E_{\rm ph} \rangle / E_r}$. Therefore, as a function of recoil energy, the free-ion approximation becomes a good one parametrically slower than the impulse approximation.

\subsubsection{Impulse approximation}
\label{subsubsec:ia}

When the energy given to the lattice is sufficiently large ($E \gg \langle E_{\text{ph}} \rangle$), it is possible to further simplify the harmonic structure factor in Eq. \ref{eq:phononexact} by using the so-called  \textit{impulse approximation} (ia)~\cite{doi:10.1119/1.1987042, schober2014introduction,Knapen:2020aky}. The integrand in Eq.~\eqref{eq:correlator} has support on a range of energies close to the typical phonon energy, $\langle E_{\rm ph} \rangle$. If the energy released to the crystal is $E$, the scattering process happens over time scales $t \sim 1 / E$. When $E \gg \langle E_{\rm ph} \rangle$ one can expand Eq.~\eqref{eq:correlator} around $t=0$, to obtain
\begin{align} \label{eq:expansion}
    f(t) = \langle E^{-1}_{\text{ph}} \rangle + i t - \frac{\langle E_{\text{ph}} \rangle t^2}{2} + \mathcal{O}\left( \langle E_{\rm ph} \rangle^2 t^3 \right) \,.
\end{align}
One can justify this procedure more rigorously through a steepest descent analysis~\cite{schober2014introduction, doi:10.1119/1.1987042, Campbell-Deem:2022fqm}.
When used in Eq.~\eqref{eq:Sharmonic}, this returns the dynamic structure factor in the impulse approximation, where it is well described by a Gaussian envelope centered around the recoil energy expected in an elastic recoil of a free nucleus, $E_r = q^2/2m_{\rm N}$. Specifically,
\begin{align}\label{eq:impulse}
    S^{\rm (ia)}({\bm q}, E) \simeq   \frac{1}{\sqrt{2 \pi \Delta^2}} e^{-\frac{(E - E_r(q))^2}{2 \Delta^2} } \,, 
\end{align}
where the width of the Gaussian is determined by $\Delta^2 \equiv E_r \langle E_{\text{ph}} \rangle $. 

The impulse approximation can be intuitively understood due to the following fact (see also our discussion in Section~\ref{sec:overview}):  
when the collision happens over times much shorter than the typical time scale characterizing the lattice dynamics (the inverse phonon energy), the final nucleus does not have time to probe the lattice potential. Its wave function can therefore be treated as a free plane wave. While this considerably simplifies the problem, it breaks down for small momentum transfers, i.e., when $q^2/2m_{\rm N} \lesssim \langle E_{\rm ph} \rangle$. For lower momenta, the full lattice potential becomes important, and phonon excitations must be properly accounted for in the dynamic structure factor.

\subsubsection{Improved impulse approximation} \label{sec:iia}

Within our simplifying assumption of an harmonic crystal, there exists an additional method to calculate the dynamic structure factor that combines the two strategies outlined in Subsections~\ref{subsubsec:recursion} and \ref{subsubsec:ia}, and allows us to extend the analytical calculation to small momentum transfers.
This procedure eases the computational burden necessary to recursively compute high orders of $T_p(E)$ (Eq.~\eqref{eq:Tprecursion}). The following result is known in the neutron scattering literature~\cite{doi:10.1119/1.1987042}, but has not yet been applied in the DM direct detection context. 
The idea is that, following the standard WKB approach, the wave function of sufficiently excited final states closely resembles a plane wave. For such states one then recovers the conditions of applicability of the impulse approximation, as explained above.
The operative procedure is then to keep the exact elastic, single phonon, and up to $p_{\text{ex}}$ multi-phonon response in Eq.~\eqref{eq:phononexact}, and to only expand $f(t)$ in small $t$ for higher order terms, i.e.
\begin{align} \label{eq:phononIA}
    \begin{split}
        \sum_{p \geq p_{\text{ex}}} \frac{1}{p!} \left(\frac{q^2}{2m_{\rm N}} f(t) \right)^p & \simeq \sum_{p \geq p_{\text{ex}}} \frac{\big(2W(\vec{q}) \big)^p}{p!}  \left(1+ i {\big\langle E^{-1}_{\text{ph}} \big\rangle}^{-1} t - {\big\langle E^{-1}_{\text{ph}} \big\rangle}^{-1} \frac{\langle E_{\text{ph}} \rangle t^2}{2} \right)^p  \\
        & \equiv  \sum_{p \geq p_{\text{ex}}} \frac{\big(2W(\vec{q}) \big)^p}{p!}  e^{xp} \,,
    \end{split}
\end{align}
where $x =  i {\big\langle E^{-1}_{\text{ph}} \big\rangle}^{-1} t - 
\frac{1}{2}\tilde{\Delta}^2 t^2 + \mathcal{O}\big({\langle E_{\rm ph} \rangle}^3 t^3\big)$, with $\tilde \Delta^2 \equiv \langle E_{\rm ph} \rangle \big\langle E_{\rm ph}^{-1}\big\rangle^{-1} - \big\langle E_{\rm ph}^{-1} \big\rangle^{-2}$.\footnote{Note that $\tilde \Delta^2 \geq 0$ is always guaranteed by the Cauchy--Schwarz inequality.}
Writing the total structure factor as $S(\bm{q},E) = \sum_p S_p(\bm q,E)$, this leads to an \textit{improved impulse approximation} (iia), such that
\begin{align} \label{eq:Spimproved}
    \begin{split}
    S_{p\geq p_{\rm ex}}^{\rm (iia)}(\bm q, E) \simeq{}& e^{-2W(\bm q)} \frac{{\big(2W(\bm q)\big)}^p}{p!} \frac{1}{\sqrt{2\pi p \tilde\Delta^2}} e^{-\frac{\left( E - p / \langle E_{\rm ph}^{-1} \rangle \right)^2}{2p\tilde \Delta^2}} \\
    \equiv{}& e^{-2W(\bm q)} \frac{{\big(2W(\bm q)\big)}^p}{p!} T^{(n,{\rm iia})}_p(E) \,,
    \end{split}
\end{align}
where we defined the normalized $T^{(n)}$ function as $ T^{(n)}_p(E) \equiv T_p(E) / \big \langle E_{\rm ph}^{-1} \big\rangle^p$. 
The total structure function can now be calculated easily up to arbitrary order by combining Eqs.~\eqref{eq:phononexact} and \eqref{eq:Spimproved}. 
\par

\begin{figure}
    \centering
    \includegraphics[width=\textwidth]{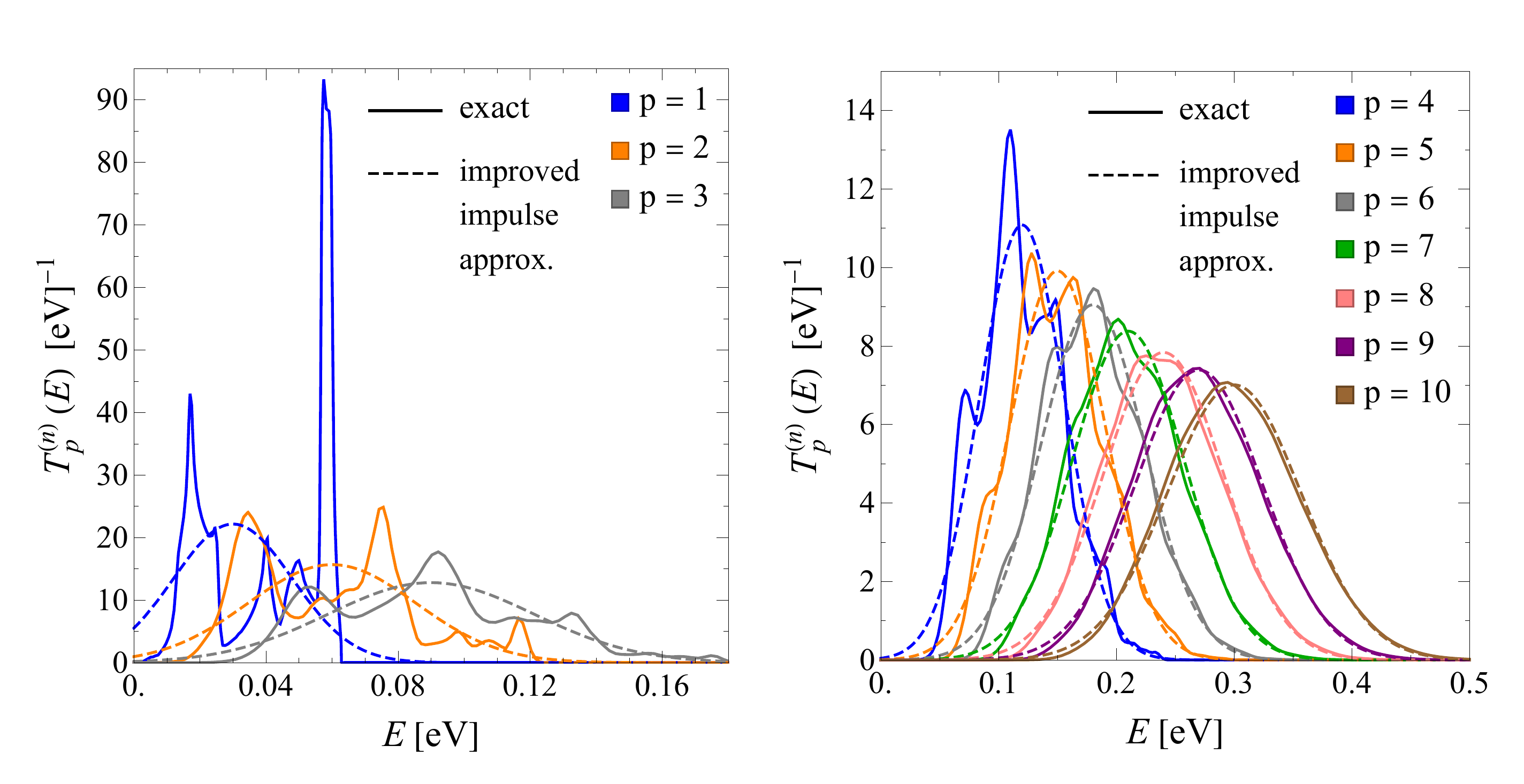}
    \caption{Comparison of the (normalized) multi-phonon terms, ${T}^{(n)}_p(E)$, obtained through the exact recursion relation in Eq.~\eqref{eq:phononexact} (\textbf{solid}) and through the improved impulse approximation (iaa) (\textbf{dashed}) for silicon. 
    The improved impulse approximation performs poorly for terms with $p \leq 3$ (\textbf{left panel}); however, already around $p \geq 4$ the approximation provides a fairly accurate description. Eventually, for $p \geq 10$ the exact multi-phonon terms no longer show any features that differ from a Gaussian, and are well described by the improved impulse approximation (\textbf{right panel}).
    }
    \label{fig:dynamic_structure_function}
\end{figure}

To compare this improved impulse approximation with the exact recursive procedure we compute the ${T}^{(n)}_p(E)$ evaluated with these two methods for the case of silicon, see Figure~\ref{fig:dynamic_structure_function}. The results highlight how the exact higher order terms calculated by the recursive procedure in Eq.~\eqref{eq:phononexact} lose sensitivity to the features of the material-specific phonon density of states. For silicon, multi-phonon terms larger than $p = 10$ no longer exhibit features distinguishing them from a simple Gaussian, and even smaller multi-phonon terms, down to $p = 4$, are decently well captured by the improved impulse approximation.\footnote{The slight difference in peak position between the improved impulse approximation and the exact result is also present in the regular impulse approximation and can be analytically accounted for~\cite{1984ZPhyB..56...13G, Liang:2022xbu}.}

The key difference between the improved impulse approximation and the standard one lies in keeping more information about the phonon density of states. While the standard approximation only depends on $\langle E_{\text{ph}} \rangle$, the procedure presented here also depends on $\big \langle E^{-1}_{\text{ph}} \big\rangle$, which enters through the width of the individual Gaussian multi-phonon terms, $\tilde{\Delta}^2$. Additionally, by keeping the exact terms up to a fixed order, the structure factor is guaranteed to be accurate for all momentum transfers, $\vec{q}$. 
In the large momentum limit the total structure function is the weighted sum of Gaussians centered around each multi-phonon term, which matches onto the regular impulse approximation.

\subsection{The dynamic structure factor in silicon }

\begin{figure}[t]
\centering
\begin{subfigure}{.5\textwidth}
    \centering
    \includegraphics[width=1\textwidth]{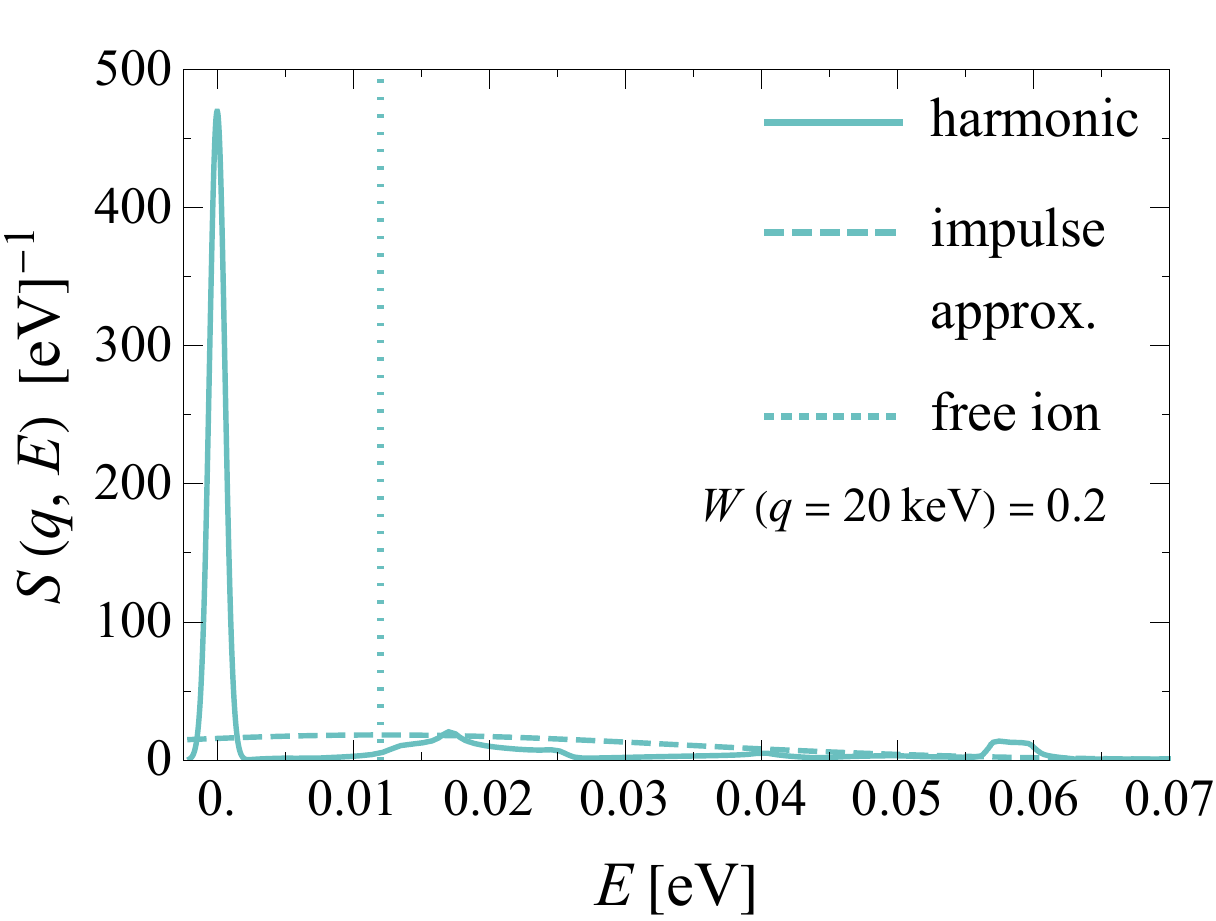}
\end{subfigure}%
\begin{subfigure}{0.5\textwidth}
    \centering
    \includegraphics[width=1\textwidth]{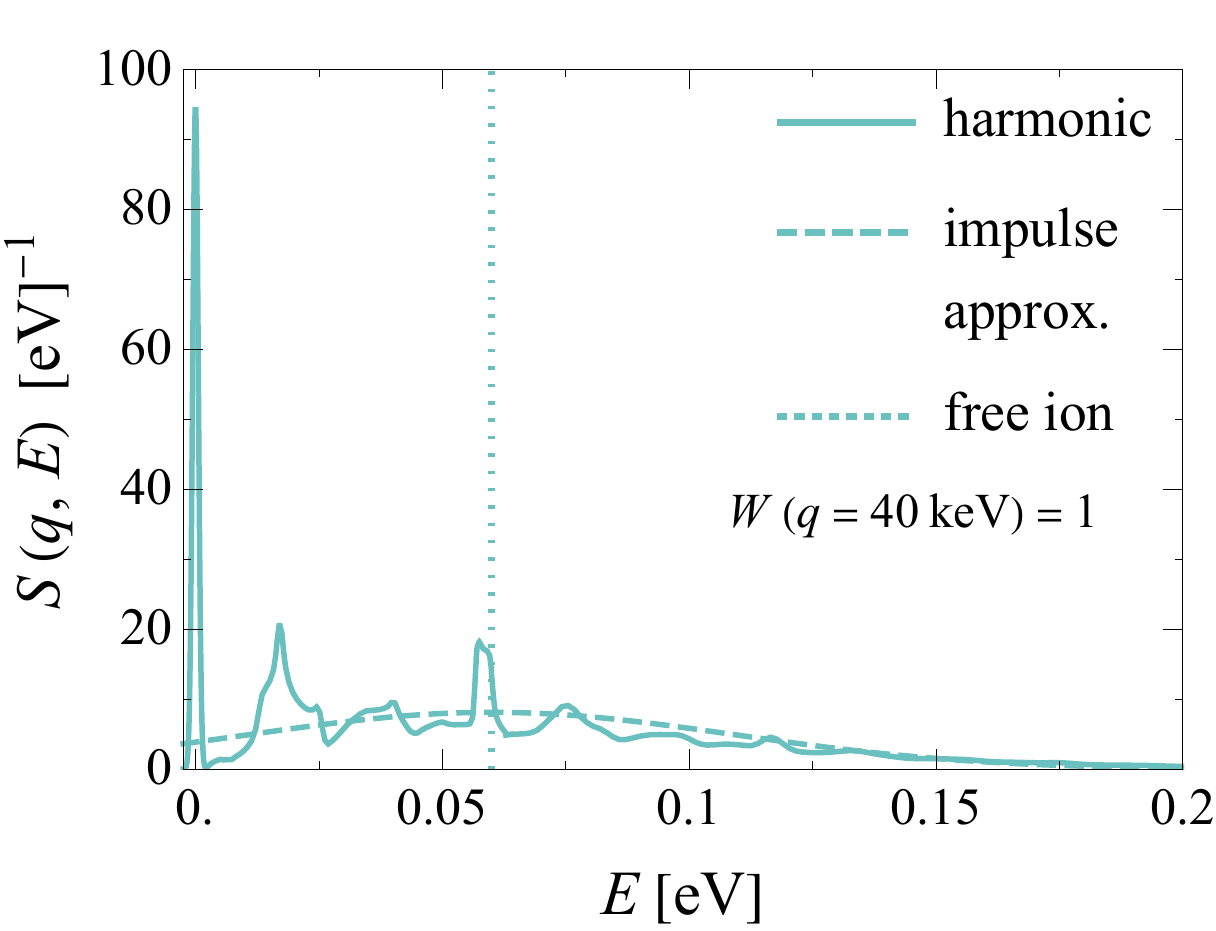}
\end{subfigure}
\begin{subfigure}{.5\textwidth}
    \centering
    \includegraphics[width=1\textwidth]{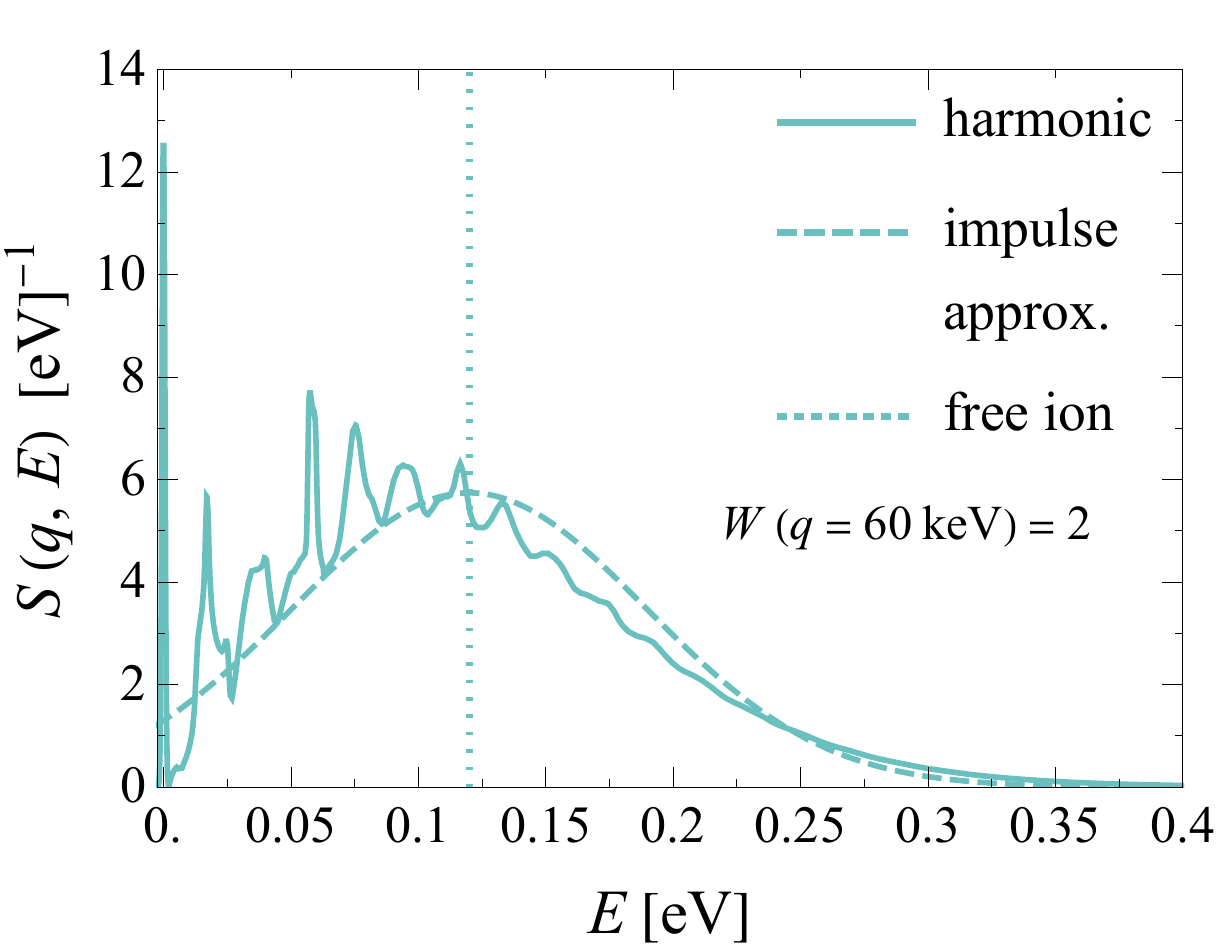}
\end{subfigure}%
\begin{subfigure}{.5\textwidth}
    \centering
    \includegraphics[width=0.98\textwidth]{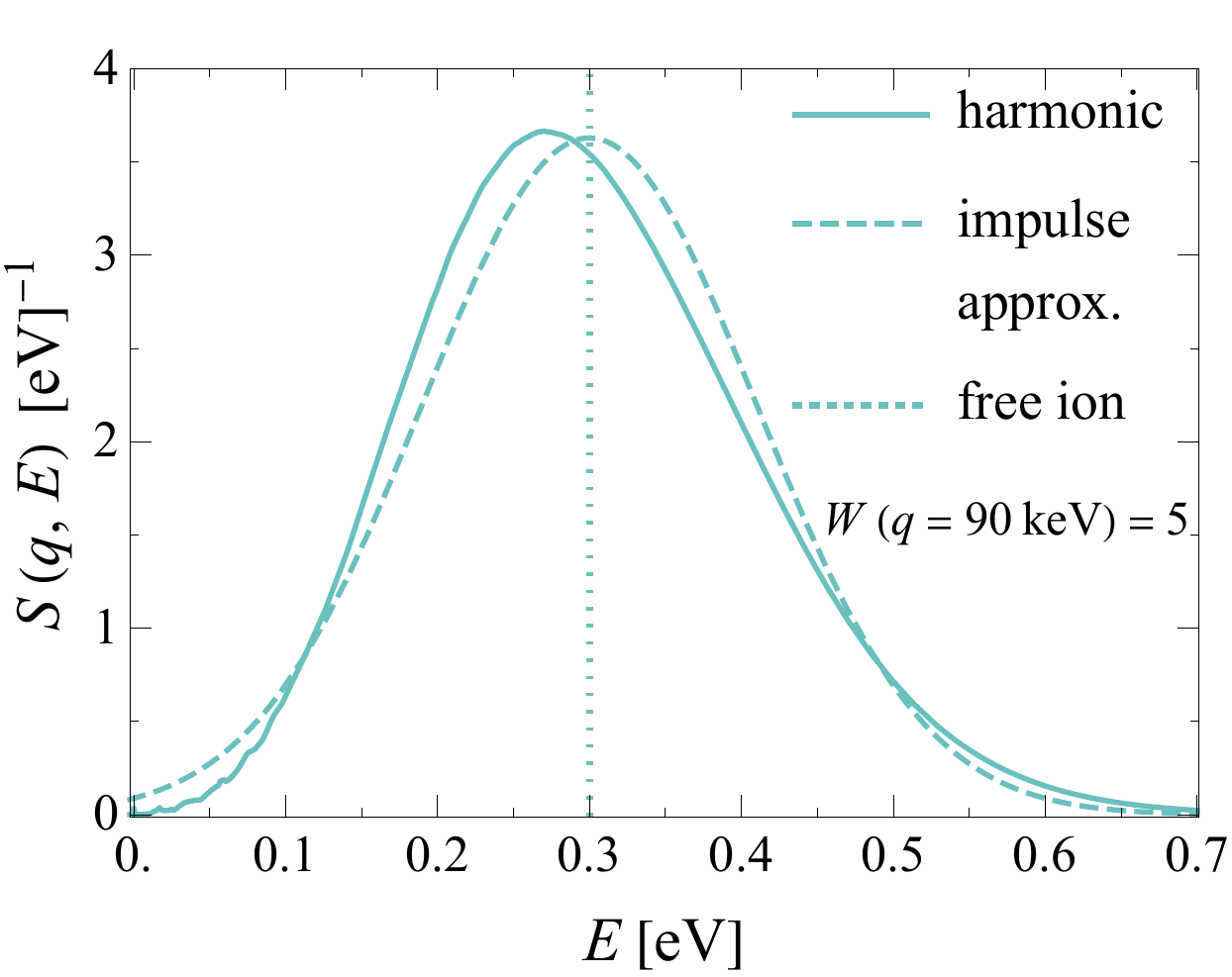}
\end{subfigure}
\caption[short]{Dynamic structure factor, $S(q,E)$, in silicon as a function of energy deposition, $E$, and for different values of the momentum transfer, $q$. We show the changes in $S(q,E)$ as the momentum transfer transitions from the small ($W(q) \ll 1$) to the large momentum ($W(q) \gg 1$) regime. To illustrate the relative contribution of the elastic term in Eq.~\eqref{eq:phononexact} we convolve it with a Gaussian envelope, $S_0(q,E) = e^{-2W(q)} \delta(E) \to e^{-2W(q)} e^{-E^2/2\delta^2} / \sqrt{2 \pi \delta^2}$, where we set $\delta^2 = 10^{-3} \tilde{\Delta}^2$. 
Note that for the bottom right figure, $q=90$~keV, the purely elastic scattering with $E=0$ is negligible, but it becomes increasingly important for lower momentum transfers.}
   \label{fig:S} 
\end{figure}

As input for our rate calculations, we evaluate $S(\bm q,E)$ in the harmonic approximation using Eq.~\eqref{eq:phononexact}, where we recursively calculate the multi-phonon terms up to order $p = 10$, and then use the improved impulse approximation for all higher order terms in the sum. The results are reported in Figure~\ref{fig:S}. As one can see, when transitioning towards small momentum transfers, the multi-phonon contribution becomes subdominant with respect to the elastic one, $S_0(q,E) = e^{-2W(q)}\delta(E)$, where the crystal recoils as a whole. Such elastic response has received less attention in the existing DM literature on the Migdal effect (although see e.g.~\cite{Kahn:2020fef,Blanco:2022pkt}). 
However, it is possible for a scattering event to lead only to ionization, without exciting any phonons in the crystal. In fact, for DM for masses $m_\chi \lesssim 50 \, \text{MeV}$ those events dominate over scattering events that are double inelastic processes, i.e., processes that produce both a Migdal ionization as well as phonons. We emphasize the importance of this contribution, noting that previous work targeting the low mass regime did not include it in their numerical evaluation of the differential rates, leading to a large discrepancy between the estimates obtained with the free-ion approximation compared to the exact result accounting for lattice dynamics~\cite{Liang:2022xbu}. The latest version of~\cite{Liang:2022xbu} now presents the corrected result.

\section{Results} \label{sec:results}

After deriving the expression for the Migdal rate within an effective field theory in Sec.~\ref{sec:fermi_theory} in terms of the dynamic structure factor, which we discuss in detail in Sec.~\ref{sec:dynamic_structure_function}, we now apply our analytical result to calculate the full and differential rates we can expect in upcoming experiments. To obtain the total rate per target mass, $R$, we divide by the detector mass $M_{\rm T}$, average over the DM's initial velocity, and multiply by the number of DM particles in the detector such that 
\begin{align} \label{eq:rate}
    R =   n_{\chi} \frac{V}{M_{\rm T}} \int d\omega \int d^3 v f_{\chi} (\vec{v})  \frac{d\Gamma(\vec{v})}{d\omega} \,,
\end{align}
where $n_{\chi}$ is the local DM number density, and $f_\chi$ is the DM's velocity distribution. 
To evaluate Eq.~\eqref{eq:rate}, we drop the off-diagonal terms, $\vec{K} \neq \vec{Q}$, in Eq.~\eqref{eq:dGammafinal} to reduce computational burden.\footnote{The combined neglected contribution is at most comparable to the diagonal terms~\cite{Kozaczuk:2020uzb}.} We also take the infinite volume limit, and replace our sums with integrals.
The  rate per target mass can then be written as,
\begin{align} \label{eq:rate2}
    R ={}& \frac{n_{\chi}}{M_{\rm T}}  \int d^3 v f_{\chi}(\vec{v}) \int d\omega \int \frac{d^3 \vec{q}}{(2 \pi)^3} \int \frac{d^3 \vec{k}_e}{(2 \pi)^3} \sum_{\bm K} 8 \pi \alpha N_{\rm T} \bigg( \frac{ g_\chi g_{\rm N}}{m_{\rm N} \omega^2} \bigg)^{\!\!2} \\
    & \times Z^2(|\vec{k}_e +\vec{K}|) \frac{(\vec{q}\!\cdot\!(\vec{k}_e + \vec{K}))^2}{\big( q^2 + m_\phi^2 \big)^2} \frac{\text{Im}\big(\!-\epsilon_{\vec{K}\vec{K}}^{-1}(\vec{k}_e,\omega)\big)}{|\vec{k}_e + \vec{K}|^2} S\left(\vec{q} - \bm{k}_e - \bm{K},\vec{q}\!\cdot\!\vec{v}- \tfrac{q^2}{2m_\chi} - \omega \right) \nonumber \,.
\end{align}
Assuming for both the velocity distribution and structure factor to be isotropic (see Appendix~\ref{app:isotropic_approximation} for details), this equation simplifies to
\begin{align}\label{eq:MigdalIonizationRate}
    \begin{split}
        R ={}& \int d\omega \frac{\alpha A^2  \bar{\sigma}_{ n}n_{\chi} N_{\rm T}}{ m_{\rm N}^2 \mu^2_{\chi n} \omega^4 M_{\rm T}}     \int  d^3 v \frac{f_{\chi}(v)}{v}\int d q \, q^3 F^2_{\text{DM}}(q) \int {d\cos} \theta_{qk}  \cos^2\theta_{qk}  \\
        & \times \int \frac{d^3 \vec{k}_e}{(2 \pi)^3} \sum_{\vec{K}} Z^2(|\vec{k}_e + \vec{K}|)  \, \text{Im}\big(\!-\epsilon_{\vec{K}\vec{K}}^{-1}(\vec{k}_e,\omega)\big) \\
        & \times \int_{0}^{E^{\text{max}}} dE \, S\left(\sqrt{q^2 + |\vec{k}_e +\vec{K}|^2 - 2 q |\vec{k}_e + \vec{K}| \cos \theta_{qk}} \, , E \right).
    \end{split}
\end{align}
In particular, we replaced the couplings with a reference cross section, $\bar{\sigma}_{ n}$, as follows: $\left(g_{\chi} g_{\rm N}  \right)^2 = \pi A^2 \bar{\sigma}_{ n} \big(q^2_0 + m^2_{\phi} \big)^2 / \mu_{\chi n}^2$, where $q_0 = m_\chi v_0$ is the typical DM momentum, $A$ is the mass number, and $\mu_{\chi n}$ the reduced DM--nucleon mass. We also introduced the form factor $F_{\rm DM}(q) \equiv \big( q_0^2 + m_\phi^2 \big)/\big(q^2 + m_\phi^2\big)$. Here we assume that the scalar mediator couples equally to protons and neutrons. One obtains the result for a vector rather than scalar mediator by replacing $m_{\phi}$ with $m_A$ in Eq.~\eqref{eq:rate2}. For a dark photon we would make the replacement  $\left(g_{\chi} g_{\rm N}  \right)^2 = \pi Z^2 \bar{\sigma}_{ p} \big(q^2_0 + m^2_{A} \big)^2 / \mu_{\chi p}^2$. 

The electron momentum appearing in the structure factor satisfies $|\bm k_e + \bm K|\lesssim 20\text{ keV}$~\cite{Knapen:2021bwg}. For sufficiently heavy DM ($m_\chi \gtrsim 10 \text{ MeV}$) one can then neglect it with respect to the momentum transfer, $q \gg |\bm k_e + \bm K|$. In this regime one can evaluate the angular integral analytically, which simplifies the rate to a factorized expression, 
\begin{align}\label{eq:MigdalIonizationRateapprox}
    R \simeq{}& \int d\omega \frac{2\alpha A^2  \bar{\sigma}_{n}n_{\chi} N_{\rm T}}{3 m_{\rm N}^2 \mu^2_{\chi n} \omega^4 M_{\rm T}}     \int_{{v}_{\text{min}}}^{{v}_{\text{max}}}  d^3 {v} \frac{f_{\chi}(v)}{v} \int \frac{d^3 \vec{k_e}}{(2 \pi)^3}  \sum_{\vec{K}} Z^2(|\vec{k}_e + \vec{K}|)  \text{Im}\big(\!-\epsilon_{\vec{K}\vec{K}}^{-1}(\vec{k}_e,\omega)\big) \notag \\
    & \times \int_{q_{\text{min}}}^{q_{\text{max}}} {d}q \, q^3 F^2_{\text{DM}}(q)  \int_{0}^{E^{\text{max}}} dE ~S(q, E) \,.
\end{align}
Near the kinematic endpoint, for low momenta when $q \sim |\vec{k}_e + \vec{K}|$, this factorization breaks down and one should evaluate the full multi-dimensional integral in Eq.~\eqref{eq:MigdalIonizationRate}.

To numerically compute Eq.~\eqref{eq:MigdalIonizationRate}, we quantify the partial dynamic structure factor as described in Subsection~\ref{sec:iia} and illustrated in Fig.~\ref{fig:S}.
For the isotropic DM velocity distribution, we assume a standard truncated Maxwell--Boltzmann distribution, boosted by the Earth velocity with respect to the galactic rest frame which we average over the angular distribution
\begin{align}
    f_{\chi}(v) = \frac{1}{4 \pi} \int d\Omega_{\vec{v}} \frac{1}{\pi^{\frac{3}{2}} v^2_0 \left( v_0 \erf(\frac{v_{\text{esc}}}{v_0}) - \frac{2 v_{\text{esc}}}{\sqrt{\pi}} e^{-\frac{v^2_{\text{esc}}}{v^2_0}} \right)} e^{-\frac{(\vec{v}+ \vec{v}_{\rm e})^2}{v^2_0}}   \Theta(v_{\text{esc}} - | \vec{v} +\vec{v}_{\rm e} |)\,, 
\end{align}
where we set $v_{\rm e} = 240 \text{ km/s}$, $v_0 = 220 \text{ km/s}$, and $v_{\text{esc}} = 500 \text{ km/s}$. We also take the DM number density to be $n_{\chi} = 0.4 \text{ cm}^{-3} \, (1 \text{ GeV}/ m_\chi)$. For the electron momentum integral we use the data tables provided in \href{https://github.com/tongylin/DarkELF} {\tt DarkELF} \cite{Knapen:2021bwg}, which specify $Z(|\vec{k_e} + \vec{K}|)$ and $\text{Im}\big(\!-\epsilon_{\vec{K}\vec{K}}^{-1}(\vec{k}_e,\omega)\big)$, calculated via time-dependent density functional theory methods with the GPAW package \cite{PhysRevB.71.035109, Enkovaara_2010} for silicon and germanium.
\begin{figure}[t]
\centering
\begin{subfigure}{.5\textwidth}
    \centering
    \includegraphics[width=1\textwidth]{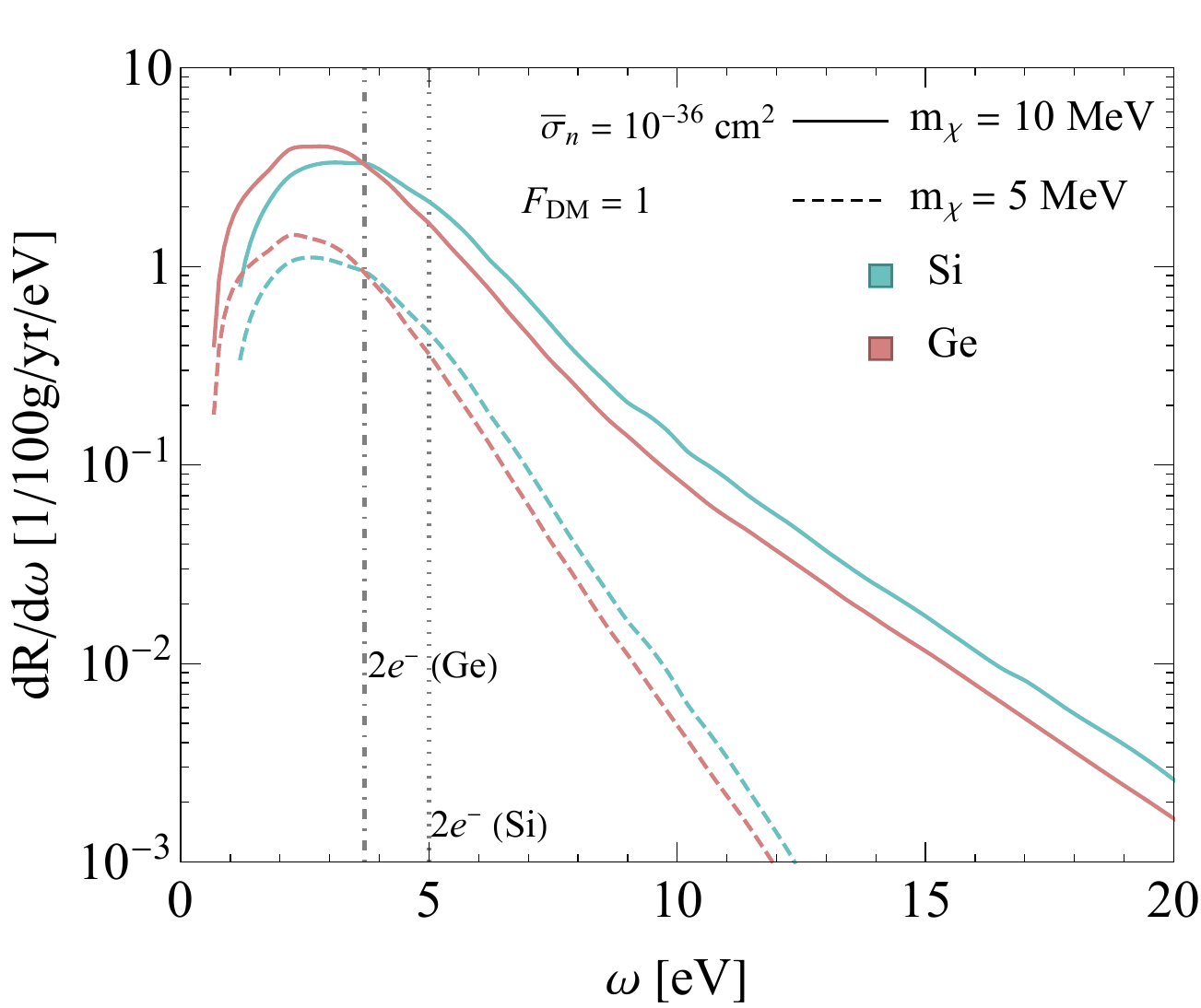}
\end{subfigure}%
\begin{subfigure}{0.5\textwidth}
    \centering
    \includegraphics[width=1\textwidth]{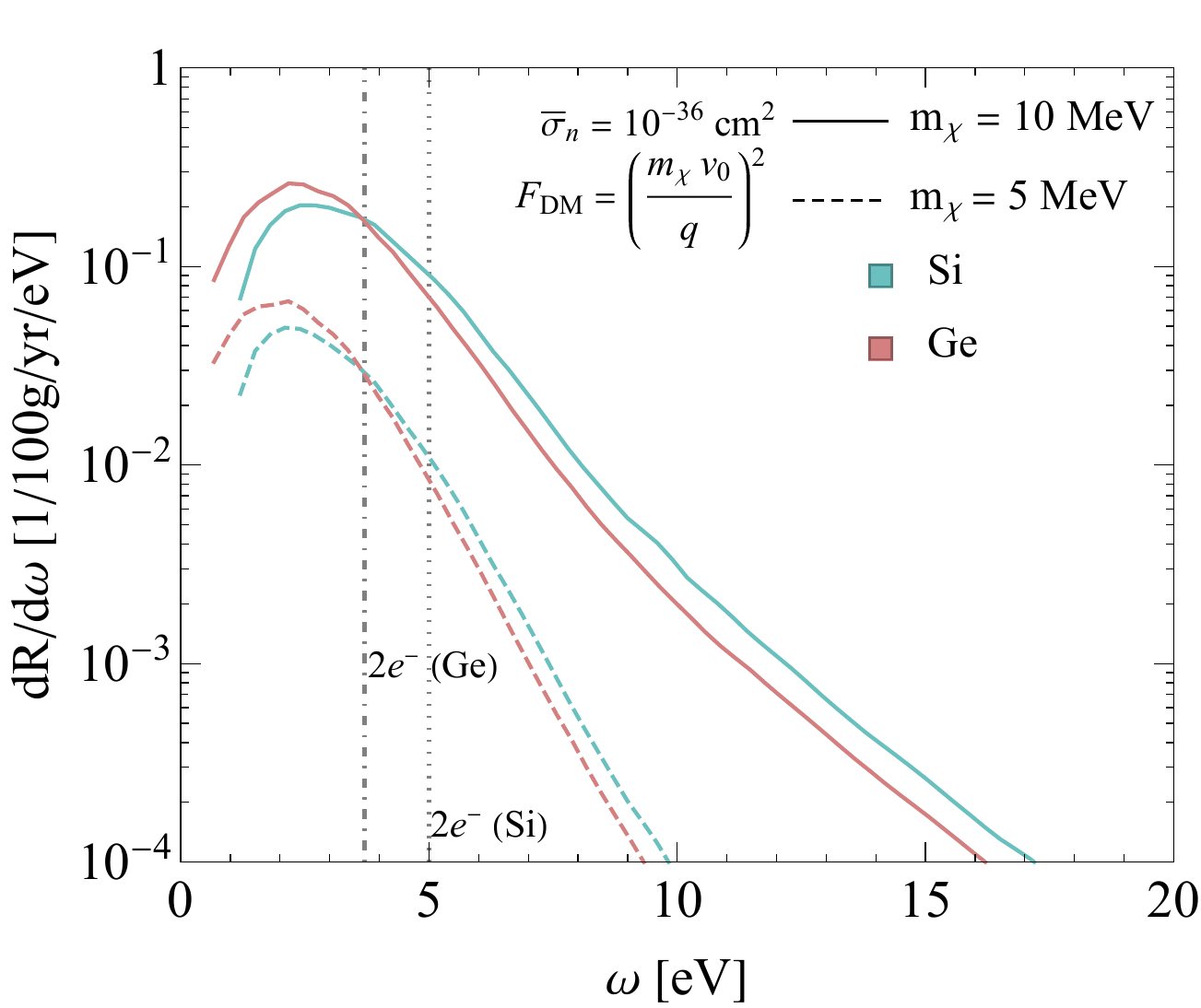}
\end{subfigure}
\caption[long]{The differential DM--nucleus Migdal scattering rate expected for 100~g of material and 1 year of exposure, computed for a  heavy (\textbf{left panel}) and a light (\textbf{right panel}) mediator, for silicon (\textbf{cyan}) and germanium (\textbf{red}), and for two different DM masses, $m_\chi = 5 \text{ MeV}$ (\textbf{dashed}), and $m_\chi = 10 \text{ MeV}$ (\textbf{solid}). The DM--nucleon reference cross-section defined below Eq.~\eqref{eq:MigdalIonizationRate} is set to $\bar \sigma_n = 10^{-36} \text{ cm}^2$. The vertical lines indicate the approximate ionization threshold required to produce two electron-hole pairs in the detector for silicon  (\textbf{dotted}, $5.0 \text{ eV}$) and germanium (\textbf{dot-dashed}, $3.7 \text{ eV}$)~\cite{Essig:2015cda}.   } \label{fig:result_dRdw} 
\end{figure}
Then we use quasi-adaptive Monte Carlo integration to evaluate the multi-dimensional integral over $\bm{k}_e$, $\cos\theta_{qk}$, $E$, $q$, $v$ and $\omega$, with the following limits of integration,
\begin{align*}
    & q_{\rm min} = m_\chi v \left( 1 - \sqrt{1 - \frac{2\omega}{m_\chi v^2}} \right) \,, \quad q_{\rm max} = m_\chi v \left( 1 + \sqrt{1 - \frac{2\omega}{m_\chi v^2}} \right) \,, \\
    & v_{\rm min} = \sqrt{2\omega / m_\chi} \,, \quad v_{\rm max} = v_e + v_{\rm esc} \,, \quad E^{\rm max} = q v - \frac{q^2}{2m_\chi} - \omega \,.
\end{align*}

Our results for the differential ionization rate, ${dR}/{d\omega}$, for silicon and germanium are shown in Fig.~\ref{fig:result_dRdw}.
The differential ionization rate peaks at energies slightly above the band gap,
and then drops with increasing $\omega$, an expected  feature due to the scaling with respect to ionization energy $dR / d\omega \propto 1 / \omega^4 $. The decrease is more pronounced for lighter DM masses as light DM has diminished phase-space available to impart large ioniziation energies. 
Experiments measure an ionization signal $Q$, the number of electron-hole pairs (denoted as $Qe^{-}$) produced in an event, rather than $\omega$ directly. We map from one onto the other using~\cite{Essig:2015cda}
\begin{equation}
Q(\omega) = 1 + \frac{(\omega - E_{\text{gap}})}{\epsilon} \,   ,
\end{equation}
where we take $\epsilon$, the mean energy per electron-hole pair, to be $3.8 \, \text{eV}$ and $2.9 \, \text{eV}$ for silicon and germanium, respectively. We assume the band gap energies $E^{\text{Si}}_{\text{gap}} = 1.2 \, \text{eV}$ and $E^{\text{Ge}}_{\text{gap}} = 0.67 \, \text{eV}$.

\begin{figure}[t!]
\centering
\begin{subfigure}{.5\textwidth}
    \centering
    \includegraphics[width=1\textwidth]{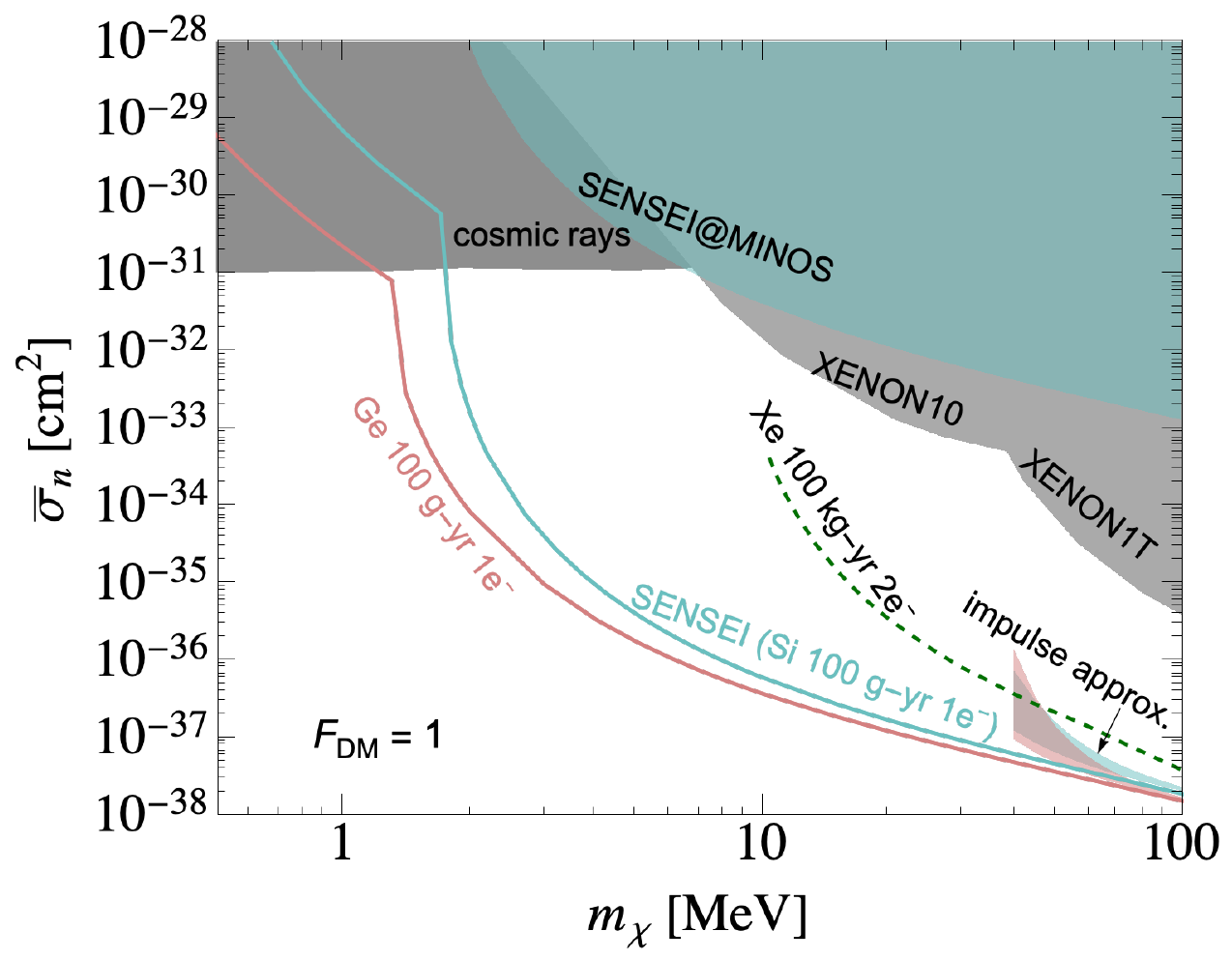}
\end{subfigure}%
\begin{subfigure}{0.5\textwidth}
    \centering
    \includegraphics[width=1\textwidth]{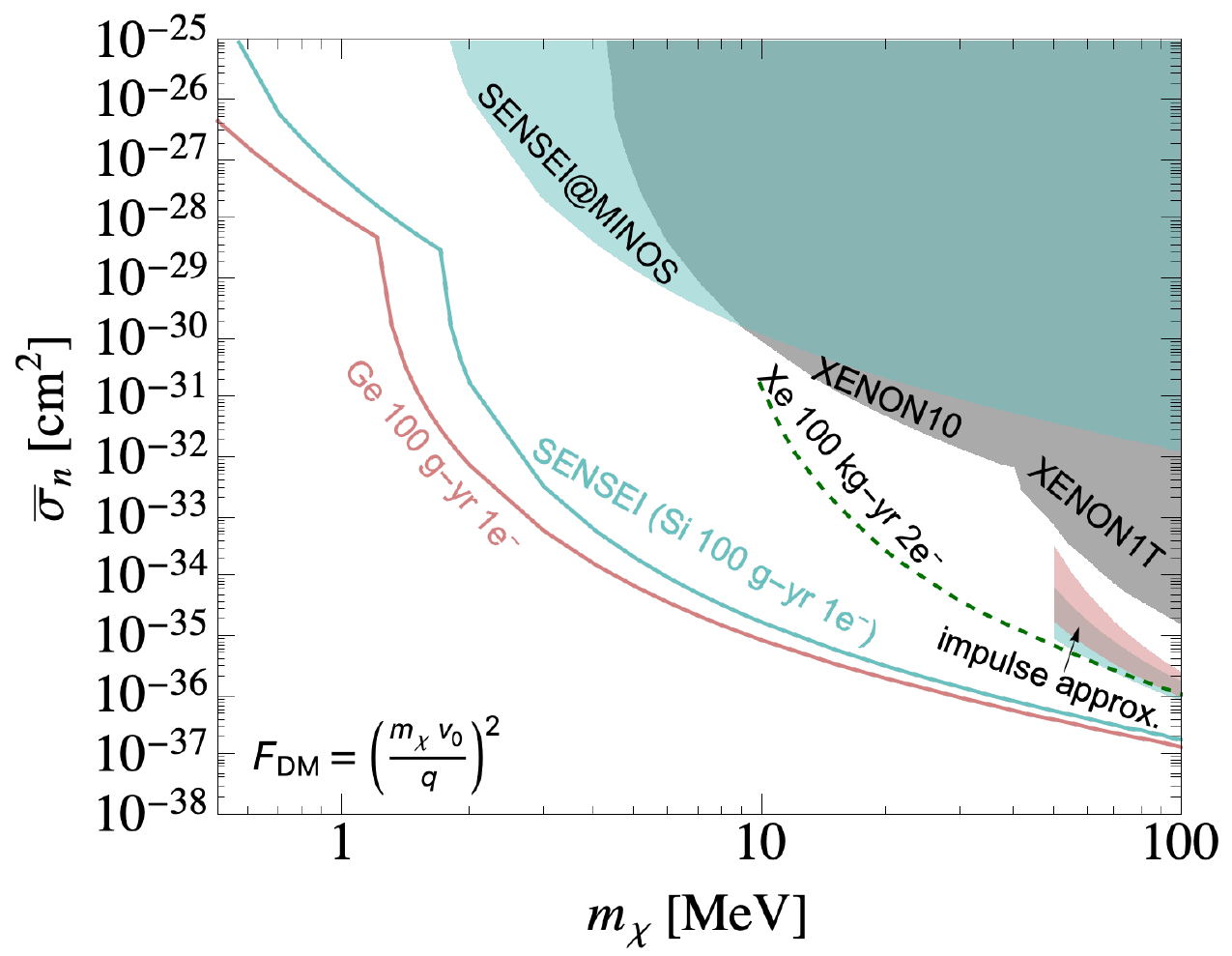}
\end{subfigure}
\caption[long]{Projections using the Migdal effect of the $90\%$ C.L.~on the DM--nucleon cross-section for a heavy (\textbf{left panel}) and light (\textbf{right panel}) mediator for SENSEI (100-g-year silicon detector) (\textbf{cyan}) and a hypothetical 100-g-year germanium detector (\textbf{red}) with $1e^{-1}$ thresholds (see text for details). We find no difference in the total rate between using the exact dynamic structure factor (Eq.~\eqref{eq:Sharmonic}) and the one obtained in the free-ion approximation (Eq. \eqref{eq:free_ion})---see text for details. 
For comparison we show projections for ``LBECA,'' a proposed 100 kg-year xenon detector, with a $2e^-$ threshold (\textbf{green, dashed})~\cite{Essig:2019xkx}. The shaded cyan and red regions correspond, respectively, to the result for silicon and germanium reported in~\cite{Knapen:2020aky}, where the authors used the impulse approximation and imposed an IR cutoff on the momentum integral when the approximation breaks down. The uncertainty band corresponds to varying the cutoff from $q_{\rm min} = 2 \sqrt{2 m_{\rm N} \langle E_{\text{ph}} \rangle }$ to $q_{\rm min} = 3 \sqrt{2 m_{\rm N} \langle E_{\text{ph}} \rangle }$. The gray-shaded regions correspond to current constraints from XENON10~\cite{PhysRevLett.110.249901}, XENON1T~\cite{PhysRevLett.123.251801}, and a recast of XENON1T data for cosmic-ray up-scattered DM~\cite{Bringmann:2018cvk}. The cyan shaded region indicates our recast of the SENSEI@MINOS~\cite{SENSEI:2020dpa} constraints.}
\label{fig:results}
\end{figure}

In Fig.~\ref{fig:results} we, instead, show projections for the exclusion limits on the DM--nucleon cross-section, $\bar\sigma_{n}$, in the low mass region for a 100-g-year detector made out of silicon and germanium for a heavy and a light mediator. The exposure corresponds to the current plans of the SENSEI detector, which consists of a silicon target. The germanium line depicts a hypothetical detector. For our projections, we assume zero background events for the bins with $\ge 2e^{-}$, but $N_{\text{bgk}} =10^5$ single-electron background events; we assume the same number of background events for SENSEI as for the hypothetical germanium detector.  This background estimate corresponds to the number of single-electron events anticipated at SENSEI for the displayed exposure.  The projections are dominated by the background free $2e^{-}$-threshold shown in Fig.~\ref{fig:result_dRdw} for masses $m_\chi \gtrsim 2 \text{ MeV}$, where we integrate the differential rate shown from the displayed $2e^-$-threshold to $\sim 20 \, \text{eV}$ to obtain the projected number of events. Our $90 \%$ C.L. corresponds to an upper limit of 2.4 events. For smaller masses, the $1e^{-}$ bin eventually delivers dominant constraints (seen as the kinks in the curves). 
We also recast the constraints from SENSEI taken at Fermilab near the MINOS hall (``SENSEI@MINOS'')~\cite{SENSEI:2020dpa}. 
Since~\cite{SENSEI:2020dpa} presented results for the $1e^-$, $2e^-$, $3e^-$, and $4e^-$ bins, we calculate the limit from each of these bins and show the best one in Fig.~\ref{fig:results}. We find that the limit is dominated by the $2e^{-}$-bin, for which we show the cross-section $\overline{\sigma}_n$ that leads to more than $9.3$ events per $2.1$ g-day exposure (we expect an analysis that combines multiple bins to yield a slightly stronger constraint).

The shaded bands in Fig.~\ref{fig:results}  correspond to the treatment in~\cite{Knapen:2020aky, Knapen:2021bwg}, where the dynamic structure factor is computed within the impulse approximation, and the momentum integral IR cutoff is varied between $q_{\rm min} = 2 \sqrt{2 m_{\rm N} \langle E_{\text{ph}} \rangle }$ and $q_{\rm min} = 3 \sqrt{2 m_{\rm N} \langle E_{\text{ph}} \rangle }$, the threshold approximating the break down of the impulse approximation. 
We find that this prescription underestimates the rate, since it artificially excludes the  low-momentum region, which is still above the kinematic threshold. The effect is more prominent for the light mediator projections, as the rate scales as $1/q$ leading to a large enhancement precisely in the small-momentum region, which then dominates the rate.\footnote{We stress that this underestimation is inherently due to imposing a cutoff on the momentum integral, rather than to the particular approximation used to describe the structure factor. We find, in fact, a similar underestimation of the full rate when using the free-ion approximation with a cutoff.}  
This cutoff-dependent rate suppression is illustrated in the right panel of Fig.~\ref{fig:results}, comparing the impulse approximation for silicon and germanium. Although germanium has a larger rate than silicon (solid lines), the uncertainty band obtained by varying the cutoff indicates a smaller rate than silicon. This is due to the cutoff $q_{\text{min}}$ being larger for the heavier germanium atom, artificially suppressing the rate.

Interestingly, we find that both the total rate, $R$, and the differential rate,  ${dR}/{d\omega}$, are insensitive to whether the dynamic structure factor is computed within the simple free-ion approximation of Eq.~\eqref{eq:free_ion}, or the more accurate harmonic one, as in Eqs.~\eqref{eq:phononexact} and \eqref{eq:Spimproved}.
We also notice that $R$ and $dR/d\omega$ are insensitive (up to sub-percent corrections) to whether one used the complete expression in Eq.~\eqref{eq:MigdalIonizationRate} or the approximate one in Eq.~\eqref{eq:MigdalIonizationRateapprox}. This is true even for small DM masses, such that $q\sim |\bm k_e + \bm K|$.
The reason for this lies in the fact that, for most of the parameter space of interest, the maximum energy that can be deposited to the crystal is much larger than the region where the structure factor has non-zero support, i.e., $E^{\rm max} \gg E_{\rm ph}, E_r$. Consequently, one can use the exact sum rule obeyed by the structure factor, $\int_0^\infty dE \, S(\bm q, E) = 1$.\footnote{One can check this analytically using, for example, the free-ion expression for the structure factor. In this case, the energy integral simply changes the limits of integration over the momentum transfer, $q_{\rm min/max} = m_\chi v \left( 1 \mp \sqrt{1-\frac{2\omega}{m_\chi v^2}} \right) \to \mu_{\chi \rm N} v \left(1 \mp \sqrt{1-\frac{2\omega}{\mu_{\chi \rm N} v^2}} \right)$. For $m_\chi \lesssim 100 \text{ MeV}$ the two coincide with very good accuracy.} Moreover, for small momentum transfer, when $E^{\rm max} \sim E_{\rm ph}$, the rate is anyway dominated by the elastic term (where the entire lattice recoils), which still satisfies the previous sum rule. Therefore, for the entire kinematic region of interest, the event rate can be written in a simple  factorized form,
\begin{equation}
\frac{dR}{d\omega}\left(\omega \right) = \frac{n_{\chi}N_T}{m_N M_T} \int^{q_{\text{max}}}_{q_{\text{min}}} dq \int d^3v v f_{\chi}(v) \frac{d\sigma}{dq}\left(q\right) \frac{dP}{d\omega}  \left(q,\omega\right) \, ,
\end{equation}
in terms of the differential cross section 
\begin{equation}
\frac{d\sigma}{dq} = \frac{A^2 \bar{\sigma}_n}{\mu^2_{\chi n} v^2 } q F^2_{\text{DM}}(q)    \, ,
\end{equation}
and the ionization or shake-off probability (see also \cite{Knapen:2020aky,Knapen:2021bwg}),
\begin{equation}
\frac{dP}{d\omega}\left(q, \omega\right) = \frac{2\alpha q^2}{3\omega^4m_N}  \int \frac{d^3 \vec{k}_e}{(2 \pi)^3}  \sum_{\vec{K}} Z^2(|\vec{k}_e + \vec{K}|)  \, \text{Im}\big(\!-\epsilon_{\vec{K}\vec{K}}^{-1}(\vec{k}_e,\omega)\big)\, .   
\end{equation}

As we just showed, the event rates inclusive in the energy released to the lattice, $E$, are insensitive to the detailed properties of the structure factor, as long as the structure factor obeys the sum rule. Extrapolating the impulse approximation $S^{(\text{ia})}(q, \omega)$ to small momentum transfers underestimates the rate by up to a factor of two since $\int_0^{\infty} dE S^{(\text{ia})}(q \to 0, \omega)  \simeq 0.5$. 

\begin{figure}[t]
\centering
\begin{subfigure}{.5\textwidth}
    \centering
    \includegraphics[width=1\textwidth]{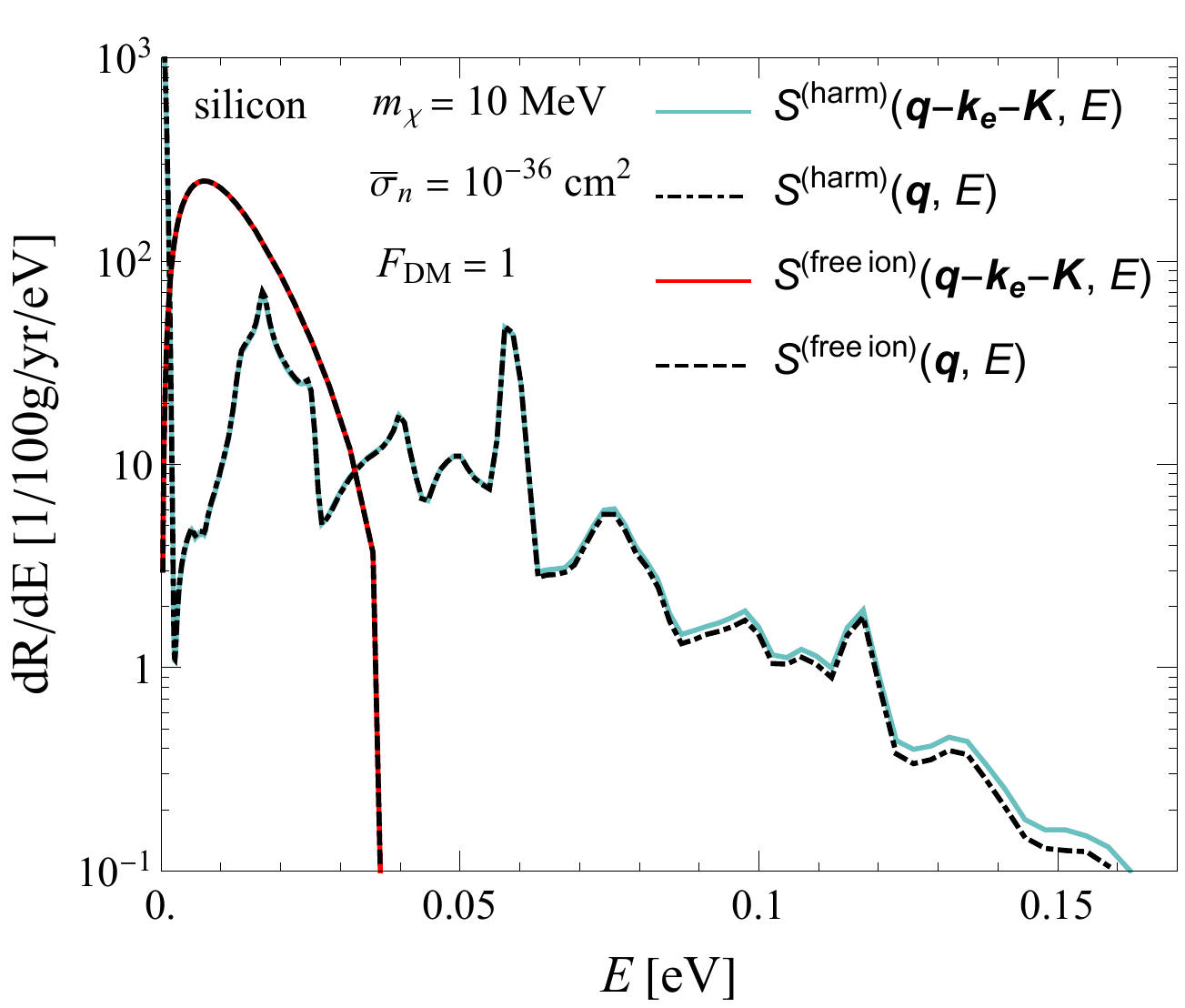}
\end{subfigure}%
\begin{subfigure}{0.5\textwidth}
    \centering
    \includegraphics[width=1\textwidth]{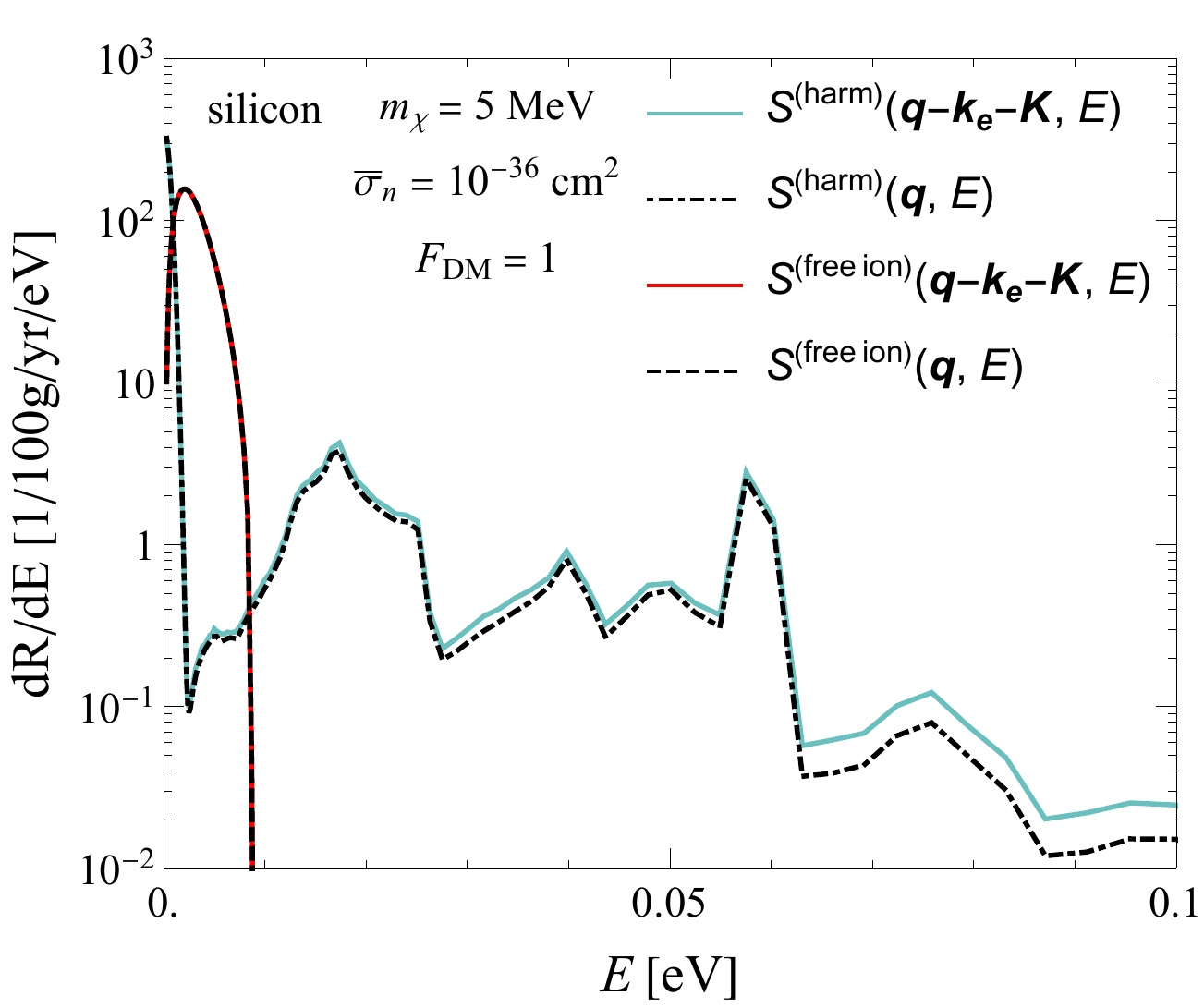}
\end{subfigure}
\caption[long]{We show the differential rate ${dR}/{dE}$, measurable by next-generation detectors, as function of energy $E$ deposited to the lattice, contrasting the harmonic dynamic structure factor with the free-ion approximation for DM with masses $m_\chi = 10 \text{ MeV}$ (\textbf{left panel}) and $m_\chi = 5 \text{ MeV}$ (\textbf{right panel}). We set the reference cross section to $\bar \sigma_n = 10^{-36} \text{ cm}^2$ and the minimum electron energy to $\omega_{\rm min} = 5 \text{ eV}$. The harmonic dynamic structure factor accurately captures the non-zero probability of phonon excitations energies above the allowed threshold in the free-ion approximation, illustrating the break-down of the free-ion approximation. Neglecting the electron momentum in the dynamic structure function results only in small differences even for the lightest DM, $m_\chi = 5 \text{ MeV}$, where the electron momentum is of order of the total momentum transfer. Consistent with Fig.~\ref{fig:dynamic_structure_function} we convolve the elastic scatterings with no energy deposition $S^{\text{(harm)}}_0 = \delta(E)$ with a Gaussian to illustrate the shift from phonon depositions to elastic scatterings off the crystal.} \label{fig:result_dRdE} 
\end{figure}

Nonetheless, our expression \eqref{eq:MigdalIonizationRate} can be used also to extract additional information about the exclusive observable of the differential energy distribution deposited to the lattice ${dR}/{dE}$, which instead strongly depends on the structure factor. Specifically,
\begin{align} \label{eq:MigdaldifferentialRateE}
    \begin{split}
        \frac{dR}{dE} &= \int d\omega \frac{2\alpha A^2 \bar{\sigma}_{n}n_{\chi} N_T}{m_N^2 \mu^2_{\chi n} \omega^4 M_T}     \int  d^3 {v} \frac{f_{\chi}(v)}{v}\int {d}q \, q^3 F^2_{\text{DM}}(q) \int^{1}_{-1}  {d\cos} \theta_{qk}  \cos^2\theta_{qk}  \\
        & \times  \int \frac{d^3 \vec{k}_e}{(2 \pi)^3}    \sum_{\vec{K}} Z^2(|\vec{k}_e + \vec{K}|)  \, 
        \text{Im}\big(\!-\epsilon_{\vec{K}\vec{K}}^{-1}(\vec{k}_e,\omega)\big)  \\
        & \times S\left(\sqrt{q^2 + |\vec{k}_e +\vec{K}|^2 - 2 q |\vec{k}_e+ \vec{K}| \cos \theta_{qk}} \, , E \right)\, \Theta(E_{\text{max}}- E) \,.
    \end{split}
\end{align}
Next-generation detectors, e.g., those employing transition edge sensor technology~\cite{SPICE}, aim at measuring phonon and multi-phonon energy depositions. Such detectors are expected to be sensitive to both the ionization as well as the lattice-energy deposition due to a DM particle hitting a nucleus in a semiconductor. 
We expect this multi-channel signal to be a powerful tool to discriminate between a DM signal and backgrounds (see also~\cite{Kahn:2020fef}). 
In Fig.~\ref{fig:result_dRdE}, we show the differential rate ${dR}/{dE}$ for silicon, for the case of a heavy mediator. 
In particular, we contrast the spectrum obtained using the free-ion approximation with the result obtained from the harmonic dynamic structure factor. The two spectra are drastically different, both in shape and in maximum energy deposits allowed, with the more accurate harmonic approximation predicting much larger ones. 
We also show that neglecting or not $|\bm k_e + \bm K|$ in the structure factor generally has no appreciable effects on the differential rates.
This is particularly advantageous, since neglecting the electron momentum in Eq.~\eqref{eq:MigdaldifferentialRateE} makes its evaluation much less computationally demanding.

\section{Summary and discussion}
\label{sec:conclusions}
When searching for sub-GeV dark matter with semiconductor targets, the distinction between high- and low-energy physics is not sharp anymore, requiring a framework that can quantify ionization $\mathcal{O}( \gtrsim  \text{eV})$ and phonon $\mathcal{O}(\sim 10 \, \text{meV})$ signals at the same time.   
A robust theoretical understanding of these detection channels is key for ensuring the continued success of experiments such as SENSEI, DAMIC-M, SuperCDMS, CDEX, and others, as well as future experiments such as Oscura~\cite{Aguilar-Arevalo:2022kqd} and SPICE~\cite{SPICE}.
In this work, we extend the regime of validity of the description of the Migdal effect in semiconductors to the lowest mass regime ($m_\chi = 1 \, - 50 \text{ MeV}$) that is kinematically able to give rise to ionization.
We accurately incorporate the bound nature of the nucleus in a crystal by taking advantage of the large separation of scales between the final electron energy and the typical energies of the excitations of the crystal lattice, formulating the problem in an effective field theory. The effective field theory allows us to write the rate for Migdal emission in a way that is universal and independent of the detailed knowledge of the inter-atomic interactions. Specifically, the vibrational modes of the crystal are all encoded in the dynamic structure factor, while the electronic dynamics is encoded in the energy loss function. Both quantities are a priori directly measurable. 
Comprehensive data quantifying the dynamic structure factor is not yet available but is obtainable through neutron scattering experiments. In the absence of data we utilize the harmonic approximation, which neglects higher order corrections to the atomic
potential, to illustrate how to model the dynamic structure function in all regimes.  

 Whether the free-ion approximation accurately captures the Migdal rate in the low mass regime had been an open question in the literature \cite{Knapen:2020aky,Liang:2020ryg,Liang:2022xbu}.
We find that the exact inclusive rate expression for the Migdal effect in a semiconductor agrees with the result one obtains by extrapolating the free-ion approximation, which we prove to be a consequence of an exact sum rule obeyed by the structure factor.
 When studying the differential rate in the energy deposited to the crystal, $dR/dE$, the above sum rule does not apply, and the harmonic structure factor predicts energy depositions to the lattice substantially larger than what is found from the na\"ive free-ion approximation.
In the future, the small energy deposits to the lattice may become an accessible observable, in which case it will be imperative to use the exact result we have derived in this work.

\begin{acknowledgments}
We are grateful to Gabriel Cuomo, Yoni Kahn, Simon Knapen, Tongyan Lin, Guido Menichetti, Aditya Parikh, and Mauro Valli for useful discussions. We thank Tien-Tien Yu and Joseph Pradler for early discussions on how to think about the Migdal effect in semiconductors in the low-momentum regime. We also thank Duncan Adam, Yoni Kahn, and Simon Knapen for comments on the manuscript. K.B.~thanks the Department of Energy's Institute for Nuclear Theory at the Universtiy of Washington for its hospitality and the Department of Energy for partial support during the completion of this work, as well as the opportunity to present and discuss with the participants of the Dark Matter in Compact Objects, Stars, and in Low Energy Experiments workshop. A.E. and R.E. are grateful to the organizers of the Pollica Summer Workshop supported by the Regione Campania, Universit\`a degli Studi di Salerno, Universit\`a degli Studi di Napoli ``Federico II'', the Physics departments ``Ettore Pancini''  and ``E R Caianiello'', and the INFN.
K.B.~acknowledges the support of NSF Award PHY1915093. 
For most of the development of this work A.E.~has been a Roger Dashen Member at the Institute for Advanced Study, whose work was also supported by the U.S.~Department of Energy, Office of Science, Office of High Energy Physics under Award No. DE-SC0009988. R.E.~acknowledges support from DoE Grant DE-SC0009854, Simons Investigator in Physics Award 623940, the US-Israel Binational Science Foundation Grant No.~2016153, and the Heising-Simons Foundation Grant No.~79921. M.S. acknowledges support from Department of Energy Grants DE-SC0009919 and DE-SC0022104. 

\end{acknowledgments}

\appendix

\section{The electron--lattice Hamiltonian} \label{app:HeL}

In this Appendix, we show how to derive the general interaction between valence electrons and the rest of the lattice, Eq.~\eqref{eq:HeL}. In our discussion, we treat core and valence electrons separately, where the former are localized around each nucleus, while the latter are not. The dielectric function is defined through the response of the material to a given electric field. Specifically, under the application of an external field, $\bm{E}_{\rm ext}$, the material responds with a macroscopic field~\cite{jackson2012classical},
\begin{align}
    \bm{E}(\bm x,\omega) = \int d^3\bm y \, \epsilon^{-1}(\bm x,\bm y,\omega) \bm{E}_{\rm ext}(\bm y,\omega) \,,
\end{align}
or, in Fourier space,\footnote{We are defining the Fourier transform of the dielectric function as $\epsilon^{-1}(\bm p, \bm q,\omega) = V^{-1} \int d^3\bm x d^3\bm y \, \epsilon^{-1}(\bm x,\bm y,\omega) e^{i \bm p \cdot \bm x - i \bm q\cdot\bm y}$.}
\begin{align} \label{eq:Ep}
    \bm E(\bm p,\omega) = \sum_{\bm q} \epsilon^{-1}(\bm p,\bm q,\omega) \bm E_{\rm ext}(\bm q,\omega) \,.
\end{align}
We work for simplicity in the isotropic approximation, where the dielectric function is just a scalar (rather than a tensor). The general case can be treated in a similar way.

The dielectric function must be invariant under discrete translations by a lattice vector, say $\bm a$: $\epsilon^{-1}(\bm x+\bm a,\bm y+\bm a,\omega) = \epsilon^{-1}(\bm x,\bm y,\omega)$. For its Fourier transform, $\epsilon^{-1}(\bm p,\bm q,\omega)$, this implies that the difference $\bm p-\bm q$ is a reciprocal lattice vector. Alternatively, one can write $\bm p = \bm k + \bm K$ and $\bm q = \bm k + \bm K^\prime$, with $\bm{K}$ and $\bm K^\prime$ belonging to the reciprocal lattice and $\bm k$ limited to the first Brillouin zone. Therefore, Eq.~\eqref{eq:Ep} can be rewritten as
\begin{align} \label{eq:EkplusK}
    \begin{split}
        \bm E(\bm k + \bm K,\omega) ={}& \sum_{\bm K^\prime} \epsilon^{-1}(\bm k + \bm K,\bm k+ \bm K^\prime,\omega) \bm{E}_{\rm ext}(\bm k + \bm K^\prime,\omega) \,.
    \end{split}
\end{align}

In order to find the Hamiltonian for a valence electron, we need to compute the electric potential generating the field above. In momentum space, electric field and potential, $\varphi$, are related by $\bm E(\bm p) = i \bm p \,\varphi(\bm p)$, and from the equation above one gets,
\begin{align}
    \begin{split}
        \varphi(\bm k + \bm K, \omega) ={}& \sum_{\bm K^\prime} \frac{(\bm k + \bm K)\cdot(\bm k + \bm K^\prime)}{{|\bm k + \bm K|}^2} \epsilon^{-1}(\bm k + \bm K,\bm k+ \bm K^\prime,\omega) \varphi_{\rm ext}(\bm k + \bm K^\prime,\omega) \\
        ={}& \sum_{\bm K^\prime} \epsilon_{\rm LL}^{-1}(\bm k + \bm K, \bm k + \bm K^\prime,\omega) \varphi_{\rm ext}(\bm k + \bm K^\prime,\omega) \,,
    \end{split}
\end{align}
where we have introduced the so-called longitudinal dielectric function,\footnote{We are grateful to Simon Knapen and Tongyan Lin for pointing this quantity out to us.} $\epsilon_{\rm LL}^{-1}(\bm p, \bm q, \omega) \equiv (\bm p\cdot\bm q/p^2) \epsilon^{-1}(\bm p, \bm q, \omega)$.
We now consider the nuclei and the core electrons as being a source of an external potential which, following standard electromagnetism, is given by
\begin{align}
    \varphi_{\rm ext}(\bm x) = \frac{1}{4\pi} \int d^3\bm y \frac{\sum_I \rho(|\bm y - \bm x_I|)}{|\bm y - \bm x|} \quad \Longrightarrow \quad \varphi_{\rm ext}(\bm p) = \sum_I \frac{\rho(p)}{p^2} e^{i \bm p \cdot \bm x_I} \,,
\end{align}
with $\rho$ the charge density, which we assume is isotropic and centered around each ion in the crystal. Introducing the symmetric dielectric matrix, $\epsilon^{-1}_{\bm K \bm K^\prime}(\bm k,\omega) \equiv \epsilon^{-1}_{\rm LL}(\bm k + \bm K,\bm k+\bm K^\prime,\omega)|\bm k + \bm K|/|\bm k + \bm K^\prime|$, the macroscopic electric potential reads
\begin{align}
    \varphi(\bm k + \bm K,\omega) = \sum_I \sum_{\bm K^\prime} \frac{\epsilon^{-1}_{\bm K \bm K^\prime}(\bm k,\omega)}{|\bm k + \bm K||\bm k + \bm K^\prime|} \rho(|\bm k + \bm K^\prime|) e^{i(\bm k + \bm K^\prime)\cdot \bm x_I} \,.
\end{align}

To obtain the Hamiltonian of a valence electron we go back to real space, and multiply the potential by the electron charge, $-e$. Defining also $\rho(k) \equiv e Z(k)$, the result is

\begin{subequations}
    \begin{align}
        H_{e \rm L} ={}& - \frac{4\pi \alpha}{V} \sum_{I} \sum_{\vec{K},{\vec{K}}^\prime} \sum_{\vec{k}}  \frac{\epsilon_{\vec{K}\vec{K}^\prime}^{-1}(\vec{k},\omega)Z(|\bm{k}+\bm{K}^\prime|)}{|\vec{k}+\vec{K}||\vec{k}+\vec{K}^\prime|} e^{i(\vec{k}+\vec{K})\cdot \vec{x}_e
        } e^{-i(\vec{k}+\vec{K}^\prime)\cdot\vec{x}_I} \,.
    \end{align}
\end{subequations}

\section{The dynamic structure factor from neutron spectroscopy}
\label{app:neutron_scattering}

We expressed the Migdal scattering rate as Eq.~\eqref{eq:dGammafinal}, which depends on the dynamic structure factor, $S(q,E)$, which incorporates the phonon degrees of freedom in the material.  
While we discussed several simplifying assumptions for $S(q,E)$ in Sec.~\ref{sec:dynamic_structure_function}, we discuss here how inelastic neutron scattering in crystals can probe $S(q,E)$. Typically, the neutron scattering cross section in a monatomic  crystal can be written as~\cite{schober2014introduction},
\begin{align}
    \frac{d^2 \sigma}{d \Omega dE} = \frac{\sigma_a}{4 \pi} \frac{k_f}{k_i}S(q,E) \,,
\end{align}
where $q$ and $E$ are the momentum and energy lost by the neutron, $k_i$ ($k_f$) is the initial (final) momentum of the neutron, and $\sigma_a$ is the total scattering cross section of a neutron with an individual atom in the crystal. Thus, a measurement of the neutron scattering rate for a monochromatic neutron beam in a crystal as a function of the final energy and the direction of the neutron can directly provide a measurement of the structure factor.

This measurement can be performed at a neutron spectroscopy experiment~\cite{HUDSON200625}. Typically, a neutron beam is incident on the target material, and the scattered neutrons are detected within a certain angular region. The energy and the momentum lost by the initial neutron can be determined from the measurement of the time-of-flight and the angular deflection caused by the scattering.

\begin{figure}[t]
    \centering
    \includegraphics[width=0.48\textwidth]{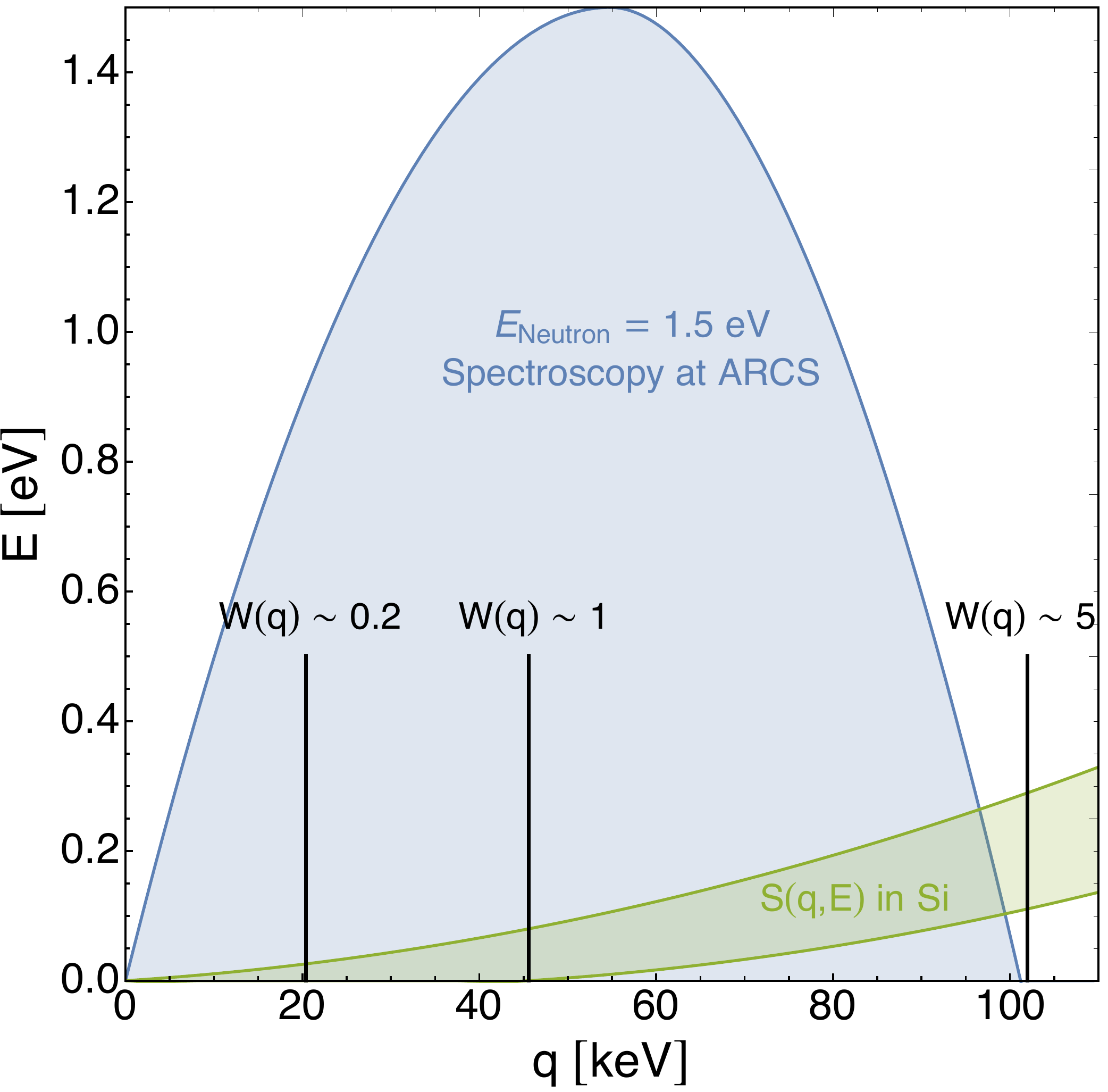}
    \caption{\textbf{Blue shaded region:} The energy-momentum phase space that can be probed kinematically with a 1.5 eV neutron beam at ARCS for a silicon sample. \textbf{Green shaded region:} The relevant energy-momentum parameter space where the dynamic structure factor in silicon is non-zero. Also shown are the momenta that correspond to the Debye--Waller factor, $W(q)$, equal to 0.2, 1, and 5.}
    \label{fig:neutron_spectroscopy}
\end{figure}

Here we consider the example of the spectrometer ARCS~\cite{abernathy2012design}. ARCS can generate a monochromatic neutron beam with typical initial energies in the range 15~meV~--~1500~meV. The angular range in the horizontal plane is from $0 \degree$ (forward scattering) to $135 \degree$ (back-scattering). Assuming perfect resolution on the final neutron energy, the energy-momentum phase space that can be probed kinematically with a 1.5~eV neutron beam at ARCS for a silicon sample is shown in Fig.~\ref{fig:neutron_spectroscopy}. We also show the region in energy-momentum space for which the dynamic structure factor in silicon is expected to be appreciably different from zero. We estimate this domain using the simple Gaussian description obtained within the impulse approximation, as in Eq.~\eqref{eq:impulse}.
We also show the momenta that correspond to the Debye--Waller factor $W$ equal to 0.2, 1, and 5, which represent the transition from the harmonic regime to the impulse regime. 

We see that with a 1.5 eV neutron, the energy-momentum parameter space where the dynamic structure factor is non-zero in silicon can be probed up to momenta of $\sim 100~\text{keV}$. Importantly, this covers the region where the Debye--Waller factor transitions from values much smaller than 1, where the dynamics are dominated by the elastic response of the entire crystal, to values much higher than 1, where the dynamics are dominated by the impulse approximation and the elastic nuclear recoil. The transition region between these two extremes is susceptible to theory uncertainties and anharmonic effects. Thus, a direct measurement of the dynamic structure factor in this region would be valuable.

\section{The isotropic approximation}
\label{app:isotropic_approximation}
In this Appendix, we derive Eq.~\eqref{eq:MigdalIonizationRate}. We start with the rate per target mass given in Eq.~\eqref{eq:rate2}, and assuming an isotropic velocity distribution $f_{\chi}(v)$, we first separate the integral over $\vec{v}$ into radial and angular components. This gives,
\begin{align} 
    \begin{split}
        R = & \frac{n_{\chi}}{M_{\rm T}}  \int  (2 \pi) v^2 dv f_{\chi}(v) \int d\omega \int \frac{d^3 \vec{q}}{(2 \pi)^3} \int \frac{d^3 \vec{k}_e}{(2 \pi)^3} \, 8 \pi \alpha N_{\rm T} \bigg( \frac{ g_\chi g_{\rm N}}{m_{\rm N} \omega^2} \bigg)^2 \\ 
        & \times \sum_{\vec{K}} Z^2(|\vec{k}_e +\vec{K}|) \frac{(\vec{q}\!\cdot\!(\vec{k}_e + \vec{K}))^2}{\big( q^2 + m_\phi^2 \big)^2} \frac{\text{Im}\big(\!-\epsilon_{\vec{K}\vec{K}}^{-1}(\vec{k}_e,\omega)\big)}{|\vec{k}_e + \vec{K}|^2} \\
        & \times \int d \cos{\theta_{vq}}~ S\left(\vec{q} - \bm{k}_e - \bm{K},qv \cos{\theta_{vq}}- \tfrac{q^2}{2m_\chi} - \omega\right) \,.
    \end{split}
\end{align}
Changing variables in the last integral to the phonon energy, $E = qv \cos{\theta_{vq}}- \frac{q^2}{2m_\chi} - \omega$, we get,
\begin{align} 
    \begin{split}
        R = & \frac{n_{\chi}}{M_{\rm T}}  \int  (2 \pi) v^2 dv f_{\chi}(v) \int d\omega \int \frac{d^3 \vec{q}}{(2 \pi)^3} \int \frac{d^3 \vec{k}_e}{(2 \pi)^3} \, 8 \pi \alpha N_{\rm T} \bigg( \frac{ g_\chi g_{\rm N}}{m_{\rm N} \omega^2} \bigg)^2 \\& \times\sum_{\vec{K}} Z^2(|\vec{k}_e +\vec{K}|) \frac{(\vec{q}\!\cdot\!(\vec{k}_e + \vec{K}))^2}{\big( q^2 + m_\phi^2 \big)^2} \frac{\text{Im}\big(\!-\epsilon_{\vec{K}\vec{K}}^{-1}(\vec{k}_e,\omega)\big)}{|\vec{k}_e + \vec{K}|^2}  \\
        & \times \frac{1}{qv}\int_{0}^{E^{\text{max}}} dE~ S(\vec{q} - \bm{k}_e - \bm{K},E) \,,
    \end{split}
\end{align}
where $E^{\text{max}}$ is the maximum possible phonon energy given by,
\begin{align}
    E^{\text{max}} =  qv- \frac{q^2}{2m_\chi} - \omega \,.
\end{align}
We then also split the integral over $\vec{q}$ into radial and angular components, and get,
\begin{subequations}
    \begin{align}
        R ={}& \frac{n_{\chi}}{M_{\rm T}}  \int (2 \pi) v^2 dv f_{\chi}(v) \int d\omega \int \frac{(2 \pi)q^2 dq}{(2 \pi)^3} \int_{-1}^{1} d \cos{\theta_{qk}} \int \frac{d^3 \vec{k}_e}{(2 \pi)^3} \, 8 \pi \alpha N_{\rm T} \bigg( \frac{ g_\chi g_{\rm N}}{m_{\rm N} \omega^2} \bigg)^2 \notag \\
        & \times \sum_{\vec{K}} Z^2(|\vec{k}_e +\vec{K}|) \frac{q^2 |\vec{k}_e + \vec{K}|^2 \cos^2{\theta_{qk}}}{\big( q^2 + m_\phi^2 \big)^2} \frac{\text{Im}\big(\!-\epsilon_{\vec{K}\vec{K}}^{-1}(\vec{k}_e,\omega)\big)}{|\vec{k}_e + \vec{K}|^2} \\ & \times \frac{1}{qv}\int_{0}^{E^{\text{max}}} dE~ S(\vec{q} - \bm{k}_e - \bm{K},E) \notag \\
        ={}& \frac{n_{\chi}}{M_{\rm T}}  \int  (2 \pi) v dv f_{\chi}(v) \int d\omega \int \frac{q^3 dq}{(2 \pi)^2} \int_{-1}^{1} d \cos{\theta_{qk}} \cos^2{\theta_{qk}} \int \frac{d^3 \vec{k}_e}{(2 \pi)^3} \, 8 \pi \alpha N_{\rm T} \bigg( \frac{ g_\chi g_{\rm N}}{m_{\rm N} \omega^2} \bigg)^2 \notag \\
        & \times \sum_{\vec{K}} Z^2(|\vec{k}_e +\vec{K}|) \frac{\text{Im}\big(\!-\epsilon_{\vec{K}\vec{K}}^{-1}(\vec{k}_e,\omega)\big)}{\big( q^2 + m_\phi^2 \big)^2} \int_{0}^{E^{\text{max}}} dE~ S(\vec{q} - \bm{k}_e - \bm{K},E) \,.
    \end{align}
\end{subequations}
Now, we replace $(g_\chi g_{\rm N})^2 /(q_0^2 +m_{\phi}^2)^2 = {\pi A^2 \bar{\sigma}_n}/{\mu_{\chi n}^2}$, where $\bar{\sigma}_n$ is the reference DM--nucleon cross section evaluated at $q_0 = m_\chi v_0$, and the momentum-dependence of the interaction is encoded in the DM form factor $F_{\text{DM}}(q)=(q_0^2 + m_\phi^2)/(q^2 + m_{\phi}^2)$. This gives 
\begin{align}
    R =& \int d\omega \frac{2\alpha A^2  \bar{\sigma}_{ n}n_{\chi} N_{\rm T}}{ m_{\rm N}^2 \mu^2_{\chi n} \omega^4 M_{\rm T}}  \int  (2 \pi) v dv f_{\chi}(v) \int d\omega \int q^3 dq~ F^2_{\text{DM}}(q) \int_{-1}^{1} d \cos{\theta_{qk}} \cos^2{\theta_{qk}} \notag \\
    & \times  \int\frac{d^3 \vec{k}_e}{(2 \pi)^3} \sum_{\vec{K}} Z^2(|\vec{k}_e +\vec{K}|) \text{Im}\big(\!-\epsilon_{\vec{K}\vec{K}}^{-1}(\vec{k}_e,\omega)\big) \int_{0}^{E^{\text{max}}} dE~ S(\vec{q} - \bm{k}_e - \bm{K},E)\,.   
\end{align}
Finally, restoring the full integral over $\vec{v}$ assuming an isotropic $f_{\chi}(v)$, and also assuming the dynamic structure factor only depends on the magnitude of the momentum argument, we get 
\begin{align}
    \begin{split}
        R ={}& \int d\omega \frac{\alpha A^2  \bar{\sigma}_{ n}n_{\chi} N_{\rm T}}{ m_{\rm N}^2 \mu^2_{\chi n} \omega^4 M_{\rm T}} \int  d^3 v \frac{f_{\chi}(v)}{v}\int {d}q \, q^3 F^2_{\text{DM}}(q) \int^{1}_{-1}  {d\cos} \theta_{qk}  \cos^2\theta_{qk}   \int \frac{d^3 \vec{k}_e}{(2 \pi)^3} \\
        & \times \sum_{\vec{K}} Z^2(|\vec{k}_e + \vec{K}|)  \text{Im}\big(\!-\epsilon_{\vec{K}\vec{K}}^{-1}(\vec{k}_e,\omega)\big) \\
        & \times \int_{0}^{E^{\text{max}}} dE \,S\left(\sqrt{q^2 + |\vec{k_e} +\vec{K}|^2 - 2 q |\vec{k_e}+ \vec{K}| \cos \theta_{qk}} \,, E\right) \,,
    \end{split}
\end{align}
which is Eq.~\eqref{eq:MigdalIonizationRate}.

\bibliographystyle{apsrev4-1}
\bibliography{Migdal}

\end{document}